\documentclass[twocolumn,journal]{IEEEtran}

\ifCLASSINFOpdf
  \usepackage[pdftex]{graphicx}
  \DeclareGraphicsExtensions{.pdf,.jpeg,.png}
\else
  \usepackage[dvips]{graphicx}
\fi

\usepackage{multirow}
\usepackage{makecell}
\usepackage{amsmath,amssymb}
\usepackage{epstopdf}
\usepackage{epsfig}
\usepackage{cite}
\usepackage{subcaption}
\usepackage{CJKutf8}
\usepackage{color}
\usepackage{multicol}
\usepackage{stfloats} 
\usepackage{graphicx}
\usepackage{float}
\usepackage{bm}

\usepackage{booktabs}

\usepackage{hyperref} 
\hypersetup{
	colorlinks=true,       
	linkcolor=red,         
	anchorcolor=black,     
	citecolor=blue,        
	urlcolor=blue,         
}
\usepackage{mfirstuc}
\usepackage[switch]{lineno} 
\usepackage[linesnumbered, ruled]{algorithm2e}

\usepackage{algpseudocode}

\UseRawInputEncoding
\begin{document}
\title{
	A Multi-Scale Spatial Attention Network for Near-field MIMO Channel Estimation
}

\author{Zhiming~Zhu,~
	Shu~Xu,
	Jiexin,~Zhang,~\IEEEmembership{Student Member,~IEEE},
	Chunguo~Li,~\IEEEmembership{Senior Member,~IEEE},
	Yongming~Huang,~\IEEEmembership{Fellow,~IEEE},
	and
	Luxi~Yang,~\IEEEmembership{Senior Member,~IEEE}
}

\maketitle

\renewcommand{\thefootnote}{}
\footnotetext{
	*The latest revised version has been accepted in \textit{IEEE Transactions on Communications}.
	
	Z.~Zhu, S.~Xu, J.~Zhang, Y.~Huang and L.~Yang are with the National Mobile Communications Research Laboratory, School of Information Science and Engineering, Southeast University, Nanjing 210096, China, and also with Purple Mountain Laboratories, Nanjing 211111, China (e-mail:\{zhuzm,~shuxu,~jiexinz,~huangym,~lxyang\}@seu.edu.cn).
	
	C.~Li is with the National Mobile Communications Research Laboratory,
	School of Information Science and Engineering, Southeast University, Nanjing 210096, China (e-mail: chunguoli@seu.edu.cn).
	}

\begin{abstract}
	The deployment of extremely large-scale array (ELAA) brings higher spectral efficiency and spatial degree of freedom, but triggers issues on near-field channel estimation.
	Existing near-field channel estimation schemes primarily exploit sparsity in the transform domain.
	However, these schemes are sensitive to the transform matrix selection and the stopping criteria.
	Inspired by the success of deep learning (DL) in far-field channel estimation, this paper proposes a novel spatial-attention-based method for reconstructing extremely large-scale MIMO (XL-MIMO) channel.
	Initially, the spatial antenna correlations of near-field channels are analyzed as an expectation over the angle-distance space, which demonstrate correlation range of an antenna element varies with its position.
	Due to the strong correlation between adjacent antenna elements, interactions of inter-subchannel are applied to describe inherent correlation of near-field channels instead of inter-element.
	Subsequently, a multi-scale spatial attention network (MsSAN) with the inter-subchannel correlation learning capabilities is proposed tailed to near-field MIMO channel estimation.
	We employ the multi-scale architecture to refine the subchannel size in MsSAN.
	Specially, we inventively introduce the sum of dot products as spatial attention (SA) instead of cross-covariance to weight subchannel features at different scales in the SA module.
	Simulation results are presented to validate the proposed MsSAN achieves remarkable the inter-subchannel correlation learning capabilities and outperforms others in terms of near-field channel reconstruction.

\end{abstract}

\renewcommand{\thefootnote}{\arabic{footnote}}

\section{introduction}
\IEEEPARstart{T}{he} evolution of the upcoming sixth-generation (6G) communication systems is driving the demand for enhanced spectral efficiency (SE) and higher data rates \cite{You2025_Next, Wang2023_6Gservery}.
Hence, the extremely large-scale antenna array (ELAA) and high signal frequency  are being considered to enable novel advancements in wireless communication systems \cite{Wang2023_6Gservery,Ray2021_Vision,Cui2023_survey}.

Despite the improvement of spatial degrees of freedom and spectral efficiency, extending massive MIMO to extremely large-scale MIMO (XL-MIMO) introduces new challenges for related technologies.
This arises from a fundamental shift in the properties of electromagnetic (EM) radiation affecting user equipment (UE) in XL-MIMO, where wireless communication can occur in the near-field region \cite{Sun2025_How,Huang2025_Structured}.
The EM radiation field is classified into near-field and far-field regions, separated by the Rayleigh distance $d_R$.
In a multiple-input single-output (MISO) scenario, the Rayleigh distance $d_R$ is given by $\frac{2D^2}{\lambda}$, where $D$ and $\lambda$ are the array aperture and signal wavelength respectively \cite{Selvan2017,Zhang2022,Wan2024_Field}.
In a MIMO scenario, the Rayleigh distance $d_R$ equals $\frac{2(D_r+D_t)^2}{\lambda}$, where $D_r$ and $D_t$ are the array apertures of transmitter and receiver respectively \cite{Lu2023_Field}.
The extension of communication from the far-field region to near-field region induces the traditional channel models based on planar wavefronts mismatch with actual channel models based on spherical wavefronts.
The radiation pattern with spherical wavefronts depends not only on the angle of arrival/departure (AoA/AoD) but also on the communication distance between the transmitter and receiver \cite{Wan2024_Field}.
Therefore, beamforming in XL-MIMO can enhance communication, localization sensing and interference management through precise spatial energy focusing \cite{Zhang2023_beamfocus_NF,Huang2024_Challenges,Shi2024_Joint,Hua2024_Secure}.
However, acquiring accurate channel state information (CSI) faces greater challenges than in massive MIMO due to the inclusion of additional distance-related information.
Moreover, existing channel estimation schemes in massive MIMO experience performance degradation in near-field scenarios.
Therefore, developing effective channel estimation schemes for XL-MIMO is of paramount importance.

Significant attention has been given to near-field channel estimation in recent years.
Due to the challenges associated with acquiring CSI in XL-MIMO, conventional approaches extensively studied in massive MIMO have been developed for near-field channel estimation.
Least squares (LS)- and minimum mean squared error (MMSE)-based estimators can be directly applied to near-field channel estimation.
However, LS is highly sensitive to noise and MMSE performance constrained by the accuracy of channel statistical information and its computational complexity. 
Actually, the superior performance of existing estimation schemes is largely attributed to the low-rank property of channel matrices in massive MIMO \cite{Xie2016_Overview}.
Far-field channel estimation is traditionally formulated as an angular sparse recovery problem to employ compressive sensing (CS) algorithms, such as the classical orthogonal matching pursuit (OMP) algorithm \cite{Lee_CEomp,Alkhateeb2014_Channel}, the approximate message passing (AMP) algorithm \cite{Guan2024_Message, Donoho_Message} and the sparse Bayesian learning (SBL) algorithm \cite{SBLoffgrid_Yang,SBL_static_time}.
Inspired by the employment of CS in channel estimation, authors of \cite{Han2020_Channel} constructed a transform matrix for near-field channels by uniformly sampling the angle-distance space into grids, so that the near-field channel can be sparse representation in the transform domain and reconstructed via CS algorithms.
Then, authors of \cite{NearCE_Dai} refine the transform matrix into a polar-based dictionary, where the distance dimension is sampled non-linearly to reduce dictionary overhead.
Benefiting from the sparsity in the polar domain, the simultaneous OMP algorithm is employed to recover effectively near-field channel.
The works in \cite{Hu2023_Hybrid,Yue2024_Hybrid} exploit the inherent sparsity in the hybrid field, where the far-field path components and near-field path components exist simultaneously.
A significant implementation considers near-field reconfigurable intelligent surface (RIS)-assisted communication systems \cite{Wu2022_Field, Chen2024_Channel, Yu2023_Channel}.
RIS-assisted channels are modeled as near-field XL-MIMO channel.
The OMP \cite{Wu2022_Field,Chen2024_Channel} and SBL \cite{Yu2023_Channel} are employed to reconstruct the polar-domain sparse channel parameters.
Meanwhile, some researchers have noticed that these CS-based schemes with polar dictionary still suffer from huge codebook overhead.
To counter this, Lu et al. \cite{Lu2024_Field} proposed the damped Newtonized OMP algorithm to reconstruct the channel in angle-distance space, utilizing a hierarchical sub-dictionary to refine the path parameters step by step.
In \cite{Near_Zhang}, the near-field channel is represented sparsely in the angle domain with parameterizing the distance to improve the effectiveness of OMP.
Additionally, \cite{zhu_twc} proposed a novel adaptive joint SBL estimation algorithm among subcarriers, where the distance information is refined iteratively with the coarse angular information.

However, the aforementioned CS-based schemes are highly sensitive to dictionary selection and the setting of stopping criteria in both far-field and near-field channels.
Therefore, machine learning (ML) techniques have gained significant attention in channel estimation tasks \cite{Wang2020_Thirty,Xie2020_Dictionary,He2018_Deep,Dong2019_CNN_CE}, which focus on mapping the channel features through intelligent data-driven analysis without assuming sparsity.
In particular, ML methods have been extensively explored for far-field channel estimation over the years.
For example, deep learning (DL) algorithms are developed to learn the channel dictionary for massive MIMO \cite{Xie2020_Dictionary}, deep unfolding is utilized to learn channel structure \cite{He2018_Deep}, deep convolutional neural network (CNN) is employed to explore the spatial-frequency correlation \cite{Dong2019_CNN_CE,Xu2024_Deep} and an attention network is applied in RIS-assisted MISO channel estimation \cite{Fan2024_Spatial}.

Although the application of ML in near-field communication systems is still in its infancy, it has demonstrated unique advantages for near-field channel estimation \cite{Yu2023_Adaptive,Zhang2023_Field,Lee2022_Intelligent}.
The behavior of electromagnetic waves in near-field region is more complex, which requires more sophisticated processing to accurately capture the spatial and geometric properties of the channels.
The authors of \cite{Yu2023_Adaptive} propose a model-assisted DL framework for near-field channel estimation, where a DL-based estimator is incorporated into the iterative procedure of the existing channel schemes.
Similarly, the near-field channel estimation task is formulated as a CS problem, addressed using the learned iterative shrinkage and thresholding algorithm (ISTA) in \cite{Zhang2023_Field}.
Subsequently, CNN is utilized to capture the spatial features among adjacent antennas and subcarriers \cite{Lee2022_Intelligent}.
In essential, CNN treats the channel reconstruction as an image denoising task, extracting local features through convolution with fixed filters and the channel matrix.
However, the non-linearity of near-field channels introduces more complex spatial correlations among channel matrix entries, while the receptive field of CNN constrains their ability to capture global contexts.
Fortunately, the in-context learning (ICL) capabilities of transformers have garnered significant attention in recent years.
Transformers leverage self-attention mechanisms to capture relationships between all elements in the input sequence, regardless of their distance \cite{Guo2023_How}.

In this paper, we concentrate on addressing near-field MIMO channel reconstruction using a transformer-based approach.
Leveraging the antenna correlation of near-field channels, we design a spatial attention (SA) module specifically tailored to capture comprehensive correlations among antennas, thereby enabling accurate channel reconstruction.
Specifically, the main contributions of this work are as follows:
\begin{itemize}
	\item We address the system model for near-field XL-MIMO channel and formulate the channel reconstruction problem as a denoising task.
	The channel antenna correlation is analyzed and defined as the expectation over the angle-distance space.
	Specifically, we derive the spatial antenna correlation expressions in both the angular domain and distance domain.
	It is observed that the correlation range of an antenna element varies with its position, unlike the fixed correlation range in far-field channels.
	This highlights the ICL capability of transformer is well-suited for capturing the diverse correlations across near-field channel elements.
	
	\item Given the strong correlations between adjacent antenna elements from antenna correlation, spatial attention weights are obtained by computing the correlation across subchannels instead of all elements.
	Hence, we propose a multi-scale spatial attention network (MsSAN) for near-field MIMO channel estimation with the inter-subchannel correlation learning capabilities.
	In MsSAN, the multi-scale architecture is employed to refine the subchannel size and the spatial attention (SA) module focuses on the correlations of inter-subchannel.
	The SA module is composed by a spatial multi-head attention (SMA) and a gated spatial feed-forward network (GSFN).
	Specially, we creatively introduce the sum of dot products as SA instead of cross-covariance to weight subchannel features at different scales in SMA and a simple attention to emphasize on contextual relationships among feature elements in GSFN.

	\item Numerical simulations demonstrate the superiority of the proposed MsSAN scheme compared to other schemes and inferior of CNN in the near-field channel estimation task.
	The ablation experiments show MsSAN incorporating a multi-scale feature extraction capability outperforms SAN with a single-scale structure.
\end{itemize}

The remainder of this paper is organized as follows.
Section \ref{section:system model} introduces the near-field XL-MIMO channel model and formulates the MIMO channel reconstruction task.
Next, Section \ref{section:antenna correlation} analyzes the spatial antenna correlation of near-field channels in both the angular and distance domains.
Subsequently, in Section \ref{section:transformer}, the detail architecture of the proposed MsSAN are presented, including the design of spatial attention module and tensor flow of networks.
Simulation results are provided in Section \ref{section:simulation}.
Finally, Section \ref{section:conclusion} concludes this paper.

\textit{Notation: }
We use the following notations throughout the paper.
$\mathbf{A}$ is a matrix; 
$\mathbf{a}$ is a vector; 
$a$ is a scalar;
the superscripts $(\cdot)^{*}$, $(\cdot)^{T}$, $(\cdot)^{H}$ and $(\cdot)^{-1}$ stand for  conjugate operator, transpose operator, conjugate transpose operator and matrix inverse, respectively;
$\|\mathbf{A} \|_{F} $ and $\operatorname{tr}(\mathbf{A})$ are the Frobenius norm and trace of $\mathbf{A} $, respectively;
$\mathbf{I}_{N}$ is the $N \times N$ identity matrix;
$[\mathbf{a}]_{i}$ and $[\mathbf{A}]_{i,j}$ denote $i$-th entry of $\mathbf{a}$ and entry at the $i$-th row and $j$-th column of $\mathbf{A}$;
$\mathbf{x}\sim \mathcal{CN}(\mathbf{a}, \mathbf{A})$ is a complex Gaussian vector with mean $\mathbf{a}$ and covariance matrix $\mathbf{A}$.
$\mathbb{E}\{\cdot \}$ is used to denote expectation;
$\mathbb{E}_{\mathbf{a}}\{\cdot\} $ is the expectation operation over $p(\mathbf{a})$;
$\otimes$ and $\odot$ denote the Kronecker product and element-wise product, respectively.

\section{System Model and Problem Formulation}\label{section:system model}
\subsection{Near-field XL-MIMO Channel Model}

\begin{figure}[t]
	\centering
	\includegraphics[width=0.48\textwidth]{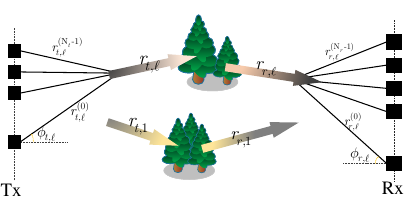}
	\caption{The near-field spherical-wave model with several scatters.}
	\label{fig:channel model}
\end{figure}

We consider a mmWave XL-MIMO uplink communication system with $N_t$-element transmit and $N_r$-element receive uniform linear arrays (ULAs), as depicted in Fig. \ref{fig:channel model}.
To highlight the near-field channel characteristics, both antenna arrays are arranged in parallel to achieve maximum phase difference \cite{Lu2023_Field}.
The speed of light and carrier frequency are denoted as $c$ and $f$.
For a uniform linear array (ULA), the antenna spacing $d$ equals the half of signal wavelength, i.e., $d=\lambda/2 $, where $\lambda = c/f$ denotes the signal wavelength.

Let $r_{r,\ell} $ denote the distance from the $\ell$-th scatter to one reference element of the receive antenna array for the $\ell$-th path component and
$r_{r,\ell}^{(n_r)}$ denotes the distance from the $\ell$-th scatter to the $n_r$-th receive element of receiver for $r_{r,\ell} $ with $n_r\in \{0,1,\cdots, N_r-1\}$.
Similarly, $r_{t,\ell} $ and $r_{t,\ell}^{(n_t)}$ represent the corresponding distances for the transmitter where $n_t\in \{0,1,\cdots, N_t-1\} $.
The physical angles $\phi_{r,\ell}$ and $\phi_{t,\ell}$ denote the AoA and AoD corresponding to $r_{r,\ell} $ and $r_{t,\ell}$, respectively.
Here, $r_{r,\ell} = r_{r,\ell}^{(0)}$ and $r_{t,\ell} = r_{t,\ell}^{(0)}$.

Let $\mathbf{H}\in \mathbb{C}^{N_r \times N_t}$ denote the XL-MIMO channel matrix from the transmitter to the receiver.
In this work, the channel between the transmitter and receiver is modeled by the parametric multipath channel model \cite{Lu2023_Field} and presented as
\begin{equation}
	\mathbf{H} = \sqrt{\frac{N_tN_r}{L}} \sum_{\ell=1}^{L} \alpha_\ell \mathbf{a}_R(\theta_{r, \ell}, r_{r,\ell}) \mathbf{a}_T^H(\theta_{t, \ell}, r_{t,\ell}),
\end{equation}
where $L $ is the number of resolved path, $L\leq \min(N_t,N_r) $, $\alpha_\ell$ is the complex gain, and the spatial angles $\theta_{r, \ell}$ and $\theta_{t, \ell}$ are sine of the physical AoA $\phi_{r,\ell}$ and AoD $\phi_{t,\ell}$ of the $\ell$-th path, i.e., $\theta_{r, \ell} = \sin\phi_{r,\ell}$ and $\theta_{t, \ell} = \sin \phi_{t,\ell}$.
The vectors $\mathbf{a}_R(\theta_{r, \ell}, r_{r,\ell}) \in \mathbb{C}^{N_r\times 1}$ and $\mathbf{a}_T(\theta_{t, \ell}, r_{t,\ell}) \in \mathbb{C}^{N_t\times 1} $ are the array response vectors at the receiver and the transmitter.
Then, the normalized array response vectors can be written as
\begin{equation}\notag
	\begin{aligned}
		&\mathbf{a}_R(\theta_{r, \ell}, r_{r,\ell}) \\&= 
		\frac{1}{\sqrt{N_r}}
		\left[e^{j\frac{2\pi}{\lambda}(r_{t,\ell}^{(0)}-r_{t,\ell}) }, \cdots,
		e^{j\frac{2\pi}{\lambda}(r_{t,\ell}^{(N_r-1)}-r_{t,\ell}) } \right]^H, 
		\\
		& \mathbf{a}_T(\theta_{t, \ell}, r_{t,\ell}) \\&=
		\frac{1}{\sqrt{N_t}}
		\left[e^{j\frac{2\pi}{\lambda}(r_{t,\ell}^{(0)}-r_{r,\ell}) }, \cdots,
		e^{j\frac{2\pi}{\lambda}(r_{t,\ell}^{(N_t-1)}-r_{t,\ell}) } \right]^H.
	\end{aligned}
\end{equation}
For the sake of notation, omit the subscript $\{r,t, \ell\}$.
Due to the spherical wavefronts, the array response vectors are functions of the spatial angle $\theta$ and distance $r$.
The distance $r^{(n)}$ of the array response vectors can be approximately expressed as \cite{zhu_twc}
\begin{equation}\label{eq:r_antenna}
	r^{(n)} \approx r  +\frac{1-\theta^2}{2r} d^2 n^2 - n d \theta.
\end{equation}
Note that the near-field range in the MISO scenario is defined as a field region within the communication distance less than the Rayleigh distance $d_R=\frac{2D^2}{\lambda}$, where $D$ is the array aperture.
For catering to the condition that largest phase discrepancy between the spherical wavefronts and the spherical wavefronts is more than $\pi/8$, the Rayleigh distance $d_R$ in the XL-MIMO scenario is calculated by $d_R = \frac{2(D_r+D_t)^2}{\lambda}$ \cite{Lu2023_Field}, where $D_r$ and $D_t$ are the receive and transmit array apertures respectively.
Therefore, MIMO extends larger near-field region, which poses greater challenges for channel recovery.

\subsection{Signal Model and Problem Formulation}
During the uplink channel estimation period, the transmitter and receiver employ $M_t$ pilot and $M_r$ beam patterns, respectively.
the transmitter employs $M_t \leq N_t$ pilot beam patterns denoted as $
\left\lbrace \mathbf{f}_p \in \mathbb{C}^{N_t\times1}: \|\mathbf{f}_p\|_2^2=1, p = 1,\cdots,M_t \right\rbrace $, and the receiver employs $M_r \leq N_r$ pilot beam patterns denoted as $
\left\lbrace \mathbf{w}_q \in \mathbb{C}^{N_r\times1}: \|\mathbf{w}_q\|_2^2=1, q = 1,\cdots,M_r \right\rbrace $.
The received vector $\mathbf{y}_p \in \mathbb{C}^{M_r \times 1}$ for the $p$-th transmit beam can be expressed as
\begin{equation}
	\mathbf{y}_p = \mathbf{W}^H\mathbf{Hf}_p s_p + \mathbf{v}_p,
\end{equation}
where $s_p $ is the transmitted pilot symbol, $\mathbf{W}=[\mathbf{w}_1, \cdots, \mathbf{w}_{M_r}] \in \mathbb{C}^{N_r \times M_r}$ represents the receive combining matrix.
The combiner output noise vector $\mathbf{v}_p = \mathbf{W}^{H}\mathbf{n}_p \in M_r \times 1$, where $\mathbf{n}_p \in \mathbb{C}^{N_r\times 1}$ is a noise vector with the noise power $\sigma^2$ following $\mathcal{CN}(0, \sigma^2 \mathbf{I}_{N_r}) $.
Collecting the all received samples $\mathbf{y}_p$ corresponding the transmit beam $\mathbf{f}_p$, the concatenated received pilot matrix  $\mathbf{Y}=[\mathbf{y}_1, \cdots, \mathbf{y}_{M_T}] \in \mathbb{C}^{M_r \times M_t}$ and is given by
\begin{equation}\label{eq:Y}
	\mathbf{Y} = \mathbf{W}^H \mathbf{HFS} + \mathbf{V},
\end{equation}
where $\mathbf{F}=[\mathbf{f}_1, \cdots, \mathbf{f}_{M_t}] \in \mathbb{C}^{N_t \times M_t}$ represents the transmit beamforming matrix and $\mathbf{V} = [\mathbf{v}_1, \cdots, \mathbf{v}_{M_t}]\in \mathbb{C}^{M_r\times M_t}$ is the concatenated noise matrix.
The matrix $\mathbf{S} \in \mathbb{C}^{M_t\times M_t}$ is a diagonal pilot symbol matrix whose diagonal entry is $\{ s_p\}$.
Without loss of generality, we assume identical pilot symbols, i.e.,  $\mathbf{S}=\sqrt{P_t}\mathbf{I}_{M_t}$, where $P_t$ is the pilot power.

In uplink channel estimation procedure, we need to reconstruct $\mathbf{H}$ with given $\mathbf{W}$ and $\mathbf{F}$.
Classically, the least-squares estimator and the linear minimum mean squared error (MMSE) estimator are widely employed to reconstruct the channel with known predesigned matrices.

\textit{1) LS Estimator:}
In vector form, \eqref{eq:Y} can be written as 
\begin{equation}
	\begin{aligned}
		\mathrm{vec}(\mathbf{Y}) &= \sqrt{P_t}(\mathbf{F}^T \otimes \mathbf{W}^H) \mathrm{vec}(\mathbf{H}) + \mathrm{vec}(\mathbf{V})\\
		&= \sqrt{P_t} \mathbf{Q}  \bar{\mathbf{h}}  + \bar{\mathbf{v}},
	\end{aligned}
\end{equation}
where $\mathbf{Q} \triangleq (\mathbf{F}^T \otimes \mathbf{W}^H) \in \mathbb{C}^{M_tM_r\times N_t N_r}$, $\bar{\mathbf{h}} = \mathrm{vec}(\mathbf{H}) $ and $ \bar{\mathbf{v}}=\mathrm{vec}(\mathbf{V})$.
Hence, the estimated channel $\hat{\mathbf{H}}_{\text{LS}}$ by LS estimator is obtained by
\begin{equation}\label{eq:ls}
	\hat{\mathbf{h}}_{\text{LS}}=\mathrm{vec}(\hat{\mathbf{H}}_{\text{LS}}) = \frac{1}{\sqrt{P_t}}(\mathbf{Q}^H\mathbf{Q})^{-1}\mathbf{Q}\mathrm{\mathbf{Y}}.
\end{equation}

\textit{2) LMMSE Estimator:}
For MMSE estimator, it seeks the weighted matrix to minimize the expected mean squared error between the true channel and the LS-based estimated channel.
The estimated channel $\hat{\mathbf{H}}_{\text{MMSE}}$ from MMSE estimator can be expressed as
\begin{equation}\label{eq:mmse}
	\mathrm{vec}(\hat{\mathbf{H}}_{\text{MMSE}}) = \mathbf{R}_{\bar{\mathbf{h}}\hat{\mathbf{h}}_{\text{LS}}}\left(\mathbf{R}_{\bar{\mathbf{h}}\bar{\mathbf{h}}} + \frac{\sigma^2}{P_t} \mathbf{I}_{N_rN_t} \right)^{-1} \hat{\mathbf{h}}_{\text{LS}},
\end{equation}
where $\mathbf{R}_{\mathbf{AB}} = \mathbb{E}\{\mathbf{AB}^H\}$ is defined as the correlation matrix between $\mathbf{A}$ and $\mathbf{B}$.
Actually, the MMSE method suffers from high computational complexity due to the involved the computation of correlation matrices and the inverse operation, especially employing the large scale antenna arrays.


\section{The spatial antenna correlation exploration}\label{section:antenna correlation}
In this section, we explore the spatial antenna correlation of the near-field channel and expound the advantages of the transformer-based neural network for spatial feature learning in XL-MIMO scenarios.

\subsection{The Channel Antenna Correlation}
Recall the array response vector and \eqref{eq:r_antenna}, the $n$-th entry of array response vector with $N$ elements for $\theta_0$ and $r_0$ is expressed as
\begin{equation}
	[\mathbf{a}(\theta_0, r_0)]_n = \frac{1}{\sqrt{N}} e^{-j\pi \left( \frac{1-\theta^2_0}{2r_0} d n^2 -  \theta_0 n \right) },
\end{equation}
where $ n=0,1,\cdots, N-1 $.
Since the near-field beams focus them on a spatial region depended on angle and path distance and each resolved path consists of several unresolvable paths in practice, each subpath arrives the antenna array plane around the mean AoA/AoD and path distance.

We set that $\phi$ is the random variable describing the AoA/AoD offset with the mean angle $\bar{\phi}_0$ and $\psi$ is the random variable describing the path distance offset with the base $\bar{r}_0$.
Then, we have $\phi_0 = \bar{\phi}_0 - \phi$ and $r_0 = \bar{r}_0 + \psi$.
The antenna correlation matrix $\mathbf{R}_{\theta, r} $ is the expectation over $(\phi, \psi)$ space, which can be expressed as
\begin{equation}
	\mathbf{R}_{\theta, r} = \mathbb{E}_{\phi, \psi}\{\mathbf{a}(\theta_0, r_0)\mathbf{a}^H(\theta_0, r_0)\},
\end{equation}
whose the $(m,n)$-th entry can be obtained by
\begin{equation}\label{eq:R_entry}
	\begin{aligned} 
	&[\mathbf{R}_{\theta, r}]_{m,n}\\ 
		=& \mathbb{E}_{\phi, \psi}\left\lbrace 
		[\mathbf{a}(\theta_0, r_0) ]_m [\mathbf{a}^*(\theta_0, r_0) ]_n
		\right\rbrace \\
		=&\mathbb{E}_{\phi, \psi } \left\lbrace 
		e^{
			-j \pi \left[
			\frac{ d (1-\theta_0^2)}{2r_0}(m^2-n^2) -(m-n)  \theta_0
			\right] 
		}
		\right\rbrace
		\\
		=&\iint_{\phi,\psi}\mathcal{P}_r(\psi) \mathcal{P}_\theta (\phi) e^{
			-j \pi \left[
			\frac{ d (1-\theta_0^2)}{2r_0}(m^2-n^2) -(m-n)  \theta_0
			\right] } \mathrm{d}{\phi} \mathrm{d}{\psi},
	\end{aligned}
\end{equation}
where $\theta_0=\sin \phi_0$. $ \mathcal{P}_\theta (\phi)$ and $\mathcal{P}_r(\psi)$ are the probability density function (PDF) of variables $\phi$ and $\psi$, respectively.
It is difficult to implement the closed-form expression of \eqref{eq:R_entry} explicitly.
As an alternative to the exploration of the correlation matrix $\mathbf{R}_{\theta, r}$, it is possible to conduct crude preliminary on how angle and distance affect near-field channels by fixing one of variables.

\subsection{The Correlation on Angular Domain}
If the distance variable is fixed, $[\mathbf{R}]_{\theta, r}$ is simplified as $[\mathbf{R}]_{\theta}$ the correlation on angular domain.
The $\mathcal{P}_\theta (\phi) $ can be interpreted as the power angle spectrum (PAS) \cite{Forenza2007_Simplified}.
Through recent measurement campaigns in indoor and outdoor, environments, it has been shown that the PAS is accurately modeled by the truncated Laplacian PDF, which is given by
\begin{equation}\label{eq:pas}
	\mathcal{P}_\theta(\phi) = \frac{\beta}{\sqrt{2} \sigma_\phi} e^{- |\sqrt{2} \phi/ \sigma_\phi|}, \quad  \phi \in [-\pi, \pi),
\end{equation}
where $\sigma_\phi$ is the standard deviation of the PAS.
$\beta = \frac{1}{1- e^{- \sqrt{2} \pi / \sigma_\phi}}$ the normalization factor.
Given the fixed distance, \eqref{eq:R_entry} can be rewritten as
\begin{equation}\label{eq:R_angle}
	[\mathbf{R}_\theta]_{m,n}= \int_{-\pi}^{\pi}\mathcal{P}_\theta(\phi) e^{
		-j \pi \left[
		\frac{ d (1-\theta_0^2)}{2r_0}(m^2-n^2) -(m-n)  \theta_0
		\right] } \mathrm{d}{\phi},
\end{equation}
By means of the trigonometric identity, we have $ \sin(\bar{\phi}_0-\phi) = \sin\bar{\phi}_0 \cos\phi - \cos\bar{\phi}_0 \sin\phi$, $\sin^2\phi_0+\cos^2\phi_0=1$ and $\cos(\bar{\phi}_0-\phi)= \cos\bar{\phi}_0\cos\phi + \sin\bar{\phi}_0\sin\phi $.
Due to high spatial resolution of the XL-MIMO, we premise that each path suffers from a small angle spreads and the PDF $\mathcal{P}_\theta(\phi)$ has a small $\sigma_\phi$. 
Hence, PAS is strongly peaking at the origin, namely, $\phi\approx 0$.
Then, expanding with first-order Taylor series
\begin{align}
	&\sin(\bar{\phi}_0-\phi) \approx \sin\bar{\phi}_0  - \phi\cos\bar{\phi}_0, \label{eq:sin} \\
	&\cos^2(\bar{\phi}_0-\phi) \approx \cos^2\bar{\phi}_0+\phi\sin2\bar{\phi}_0. \label{eq:cos}
\end{align}
Substituting \eqref{eq:sin} and \eqref{eq:cos} into \eqref{eq:R_angle}, we get
\begin{equation}\label{eq:R_angle1}
	\begin{aligned}
		&[\mathbf{R}_\theta]_{m,n}\approx 
		e^{	-j \pi \left[\frac{ d (1-\bar{\theta}_0)}{2r_0}(m^2-n^2) -(m-n)  \bar{\theta}_0
			\right] }	\\	
		&\cdot \int_{-\pi}^{\pi}\mathcal{P}_\theta(\phi) e^{-j \pi \left[
			\frac{ d(m^2-n^2) \sin2\bar{\phi}_0 }{2r_0} \phi - (m-n)\cos\bar{\phi}_0 \cdot \phi
			\right] } \mathrm{d}{\phi},
	\end{aligned}
\end{equation}
where $\bar{\theta}_0 = \sin\bar{\phi}_0$.
Equation \eqref{eq:R_angle1} consists of the product of a complex coefficient term and an integral term.
For readability, let spatial phase coefficient $\omega(m,n,\bar{\phi}_0) \triangleq  \pi(m-n)\cos\bar{\phi}_0 -\frac{ \pi d(m^2-n^2) \sin2\bar{\phi}_0 }{2r_0} $, the integral term of \eqref{eq:R_angle1} can be expressed as
\begin{equation}\label{eq:B_angle}
	\begin{aligned}\notag
		\left[ \mathbf{B}_\theta(\bar{\phi}_0)  \right] _{m,n} = \int_{-\pi}^{\pi} \frac{\beta}{\sqrt{2} \sigma_\phi} e^{- |\sqrt{2} \phi/ \sigma_\phi|} \cdot e^{j  \omega(m,n,\bar{\phi}_0)\phi } \mathrm{d}{\phi}
		\\
		= \frac{\sqrt{2}\beta}{ \sigma_\phi} \int_{0}^{\pi}  e^{- \sqrt{2} \phi/ \sigma_\phi} \cdot \cos(\omega(m,n,\bar{\phi}_0)\phi) \mathrm{d}{\phi},
	\end{aligned}
\end{equation}
where the second equality comes from the fact that the Euler's formula and the integral property of odd/even functions about symmetric bounds.
Utilizing integration by parts, the derivation of $[\mathbf{B}_\theta(\bar{\phi}_0)]_{m,n}$ is given by \eqref{eq:B_angle_res}, as shown at the top of this page.
\begin{figure*}[t]
	\begin{equation}\label{eq:B_angle_res}
	[\mathbf{B}_\theta(\bar{\phi}_0)]_{m,n}
	= \frac{\sqrt{2} \sigma_\phi \beta}{ 2+\sigma_\phi \omega^2(m,n,\bar{\phi}_0)} \left( 
	e^{-\frac{\sqrt{2}}{\sigma_\phi}\pi}\left( -\frac{\sqrt{2}}{\sigma_\phi}\cos (\pi\omega(m,n,\bar{\phi}_0)) + \omega \sin (\pi\omega(m,n,\bar{\phi}_0)) \right) 
	+\frac{\sqrt{2}}{\sigma_\phi}  
	\right)
	\end{equation}
	\hrulefill
\end{figure*}
Therefore, the closed-form of the antenna correlation on angular domain cross all the array elements is approximatively calculated by
\begin{equation}\label{eq:R_angle_res}
	[\mathbf{R}_\theta]_{m,n}\approx 
	e^{	j \pi \left[ (m-n)  \bar{\theta}_0 -\frac{ d (1-\bar{\theta}_0^2)}{2r_0}(m^2-n^2) 	\right] } \cdot [\mathbf{B}_\theta(\bar{\phi}_0)]_{m,n}.
\end{equation}
Notice that $[\mathbf{B}_\theta(\bar{\phi}_0)]_{m,n}$ is a real form and represents the magnitude of the correlation, namely, $|[\mathbf{R}_\theta]_{m,n}| = [\mathbf{B}_\theta(\bar{\phi}_0)]_{m,n}$.

\begin{figure}[t]
	\centering
	\includegraphics[width=0.35\textwidth]{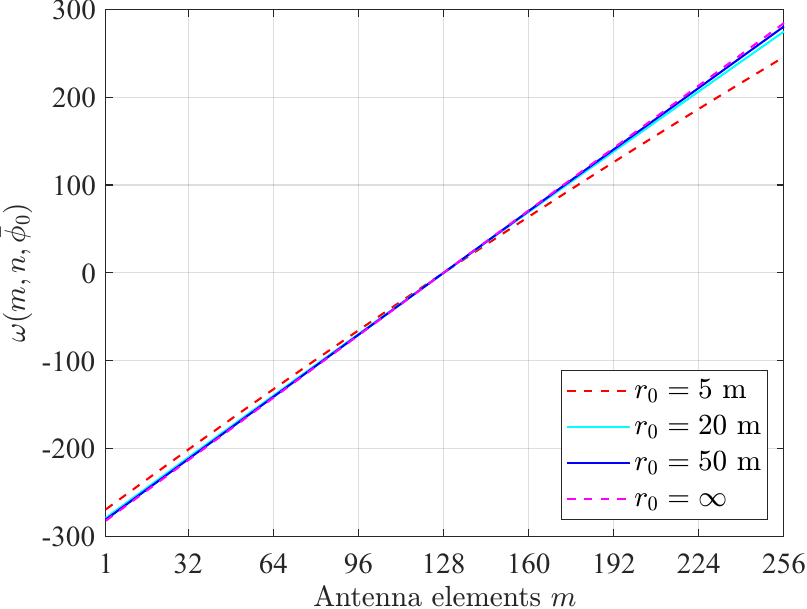}
	\caption{The spatial phase coefficient $\omega(m,n,\bar{\phi}_0)$ cross antenna elements corresponding to the center of antenna array, where $n=128$, carrier frequency $f_c = 60$ GHz and $\bar{\phi}_0= \pi/4$.}
	\label{fig:theta}	
\end{figure}

\textit{Remark 1:}
Equation \eqref{eq:R_angle_res} indicates the spatial antenna correlation of the near-field channel corresponding to the angle domain.
Especially, the correlation $[\mathbf{R}_\theta]_{m,n} $ of the far-field channel can be simplified as $e^{j \pi (m-n)  \bar{\theta}_0 } \cdot [\mathbf{B}_\theta(\bar{\phi}_0)]_{m,n} $, where $\omega(m,n,\bar{\phi}_0) =  \pi(m-n)\cos\bar{\phi}_0 $, independent of the distance.
Then, it is found that the impact of $r_0$ on $\mathbf{R}_\theta$ is fine from the simulation experiments since $\frac{ \pi d(m^2-n^2) \sin2\bar{\phi}_0 }{2r_0}\approx 0$ causes the mapping of $\omega(m,n,\bar{\phi}_0)$ in the complex space
 mainly depends on $\pi(m-n)\cos\bar{\phi}_0$.
This phenomenon results in the near-field and far-field channel almost sharing the same $|\mathbf{R}_\theta|$ corresponding to the angle domain.
This is why some algorithms implement XL-MIMO channel estimation with the orthogonal characteristics of spatial angle domain similar to far-field channel estimation \cite{zhu_twc,NearCE_Dai,Near_Zhang}.

Here, we recall the spatial phase coefficient $\omega(m,n,\bar{\phi}_0)$ presents the potential discrepancy between the near-field and far-field channels.
We present the spatial phase coefficient cross antenna elements corresponding to the center of antenna array,  as shown in Fig. \ref{fig:theta}.
The spatial phase coefficient near the center is almost equivalent to that of the far-field, while that away from the center element exists an explicit deviation compared to the far-field's.
Besides, the offset between the two sides of the antenna and the blue line is not equal, and the larger antenna size, the more pronounced the deviation of $\omega(m,n,\bar{\phi}_0)$ from the far-field's.
Namely, there is a greater difference in properties between a nearby antenna and a distant antenna.

\begin{figure}[t]
	\centering
	\includegraphics[width=0.35\textwidth]{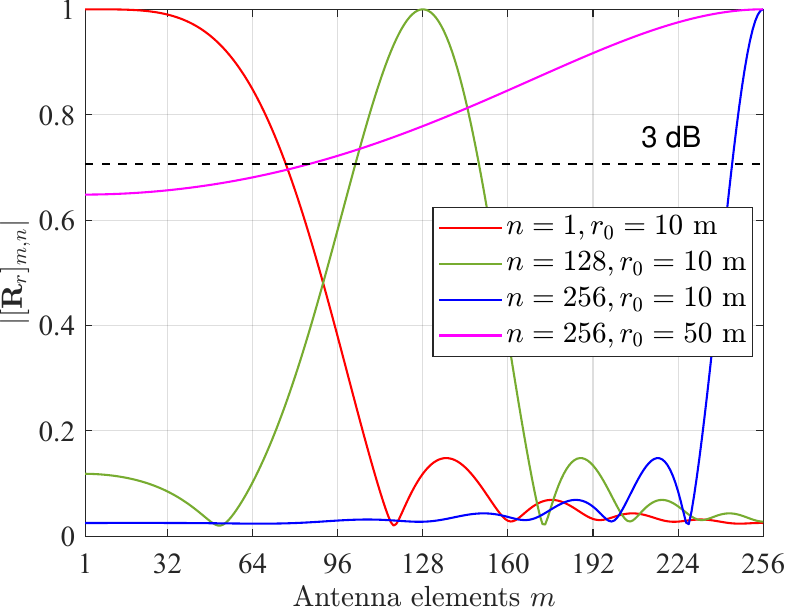}
	\caption{The antenna correlation $|\left[\mathbf{R}_r \right] _{m,n}|$ for the fixed angle, where carrier frequency $f_c = 60$ GHz, $\bar{\phi}_0= \pi/6$ and $\sigma_\psi=10$.}
	\label{fig:distance}	
\end{figure}

\subsection{The Correlation on Distance Domain}
If the angle variable is fixed, $[\mathbf{R}]_{\theta, r}$ is simplified as $[\mathbf{R}]_{r}$ the correlation on distance domain.
Given the fixed angle, \eqref{eq:R_entry} can be rewritten as 
\begin{equation}\label{eq:R_r}
	[\mathbf{R}_r]_{m,n}= \int_{0}^{\infty}\mathcal{P}_r(\psi) e^{
		-j \pi \left[
		\frac{ d (1-\theta_0^2)}{2(\bar{r}_0+\psi)}(m^2-n^2) -(m-n)  \theta_0
		\right] } \mathrm{d}{\psi}.
\end{equation}
To the best of our knowledge, the distribution $\mathcal{P}(\psi)$ has been not identified directly to characterize the change with respect to distance.
However, time and distance are inherently coupled in their relationship.
In multi-path scenario, the power delay profile (PDP) is one of the key tools for describing the distribution of signal power across different delay times.
Considering that the channel is modeled as Saleh-Valenzuela model, the power of the rays decays exponentially.
The PDP that follows an exponential distribution can be written as
\begin{equation}\label{eq:pdp}
	\mathcal{P}_t(\tau) = K_t \exp(-\tau/\sigma_\tau),\quad \tau >0,
\end{equation}
where $K_t$ is a normalization factor such that the PDP integrates to unity, $\tau$ is the delay of the subpath relative to the main path and $\sigma_\tau$ is the standard deviation of the delays.
Involving $\tau = \psi / c $, $\mathcal{P}_r(\psi)$ can be expressed by
\begin{equation}\label{eq:pdp_r}
	\mathcal{P}_r(\psi) = K_r \exp(-\psi/\sigma_\psi),\quad \psi >0,
\end{equation}
where $K_r$ is a normalization factor such that $\mathcal{P}_r(\psi)$ integrates to unity, $\psi$ is the offset distance of the subpath relative to the main path and $\sigma_\psi$ is the standard deviation of the offset.
Therefore, \eqref{eq:R_r} can be expressed as
\begin{equation}\label{eq:R_r1}
	\begin{aligned}
		\left[\mathbf{R}_r \right] _{m,n}= K_r e^{j\pi (m-n)  \theta_0 } \cdot \left[ \mathbf{B}_r(\bar{r}_0) \right] _{m,n},
	\end{aligned}
\end{equation}
where
\begin{equation}\label{eq:B_r}
	\left[ \mathbf{B}_r(\bar{r}_0) \right] _{m,n} = \int_{0}^{\infty} e^{-\frac{\psi}{\sigma_\psi}} \cdot e^{	-j \pi \frac{ d (1-\theta_0^2)}{2(\bar{r}_0+\psi)}(m^2-n^2) } \mathrm{d}{\psi}.
\end{equation}
Actually, it is difficult to directly obtain a closed-form solution of \eqref{eq:B_r}, we explore the relationship between $\mathbf{R}_r$ and $r$ using numerical simulations, as shown in Fig. \ref{fig:distance}.
As the main path $r_0$ goes to infinity, $|\left[\mathbf{R}_r \right] _{m,n}|$ converges to one.
Namely, the impact of distance on the steering vector is evident in the near-field region.
As the antenna $n$ increases, the corresponding 3dB bandwidth of $|\left[\mathbf{R}_r \right] _{m,n}|$ gradually decreases.
Besides, the different distance variables bring different $\mathbf{R}_r$ between the antenna element and adjacent antennas or distant antennas.
In summary, the non-linearity of the near-field channel phase causes the number of strongly correlated elements surrounding each channel element to be variable.
Besides, adjacent antenna elements exhibit strong correlation.
Namely, if the antenna array is divided into multiple sub-arrays, the correlation of inter sub-arrays can describe the correlation of the near-field channel.
That is the reason of applying the subchannels in \cite{Near_Zhang} and \cite{Cui2024_Field}.

\section{Transformer-based channel inference}\label{section:transformer}
In this section, we introduce an attention-based framework with the inter-subchannel correlation learning capabilities for near-field MIMO channel estimation.
The detail of the proposed MsSAN is elaborated, included the spatial multi-head attention (SMA) and the gated spatial feed-forward network (GSFN).
Especially, the spatial attention is designed to concentrate on the inter-subchannel correlation for near-field channels.

\subsection{Neural Network Framework for Channel Estimation}
Given that the DL-based channel inference schemes, such as CNN, the channel estimation problem is formulated as a denoising problem involving the coarsely estimated prior channel information.
In our framework, the noisy  $\hat{\mathbf{H}}_{\text{LS}}$ from the LS estimator are employed as the input of the deep networks.
Thus, the channel estimation network procedure $\mathcal{F}$ that is well-trained can be expressed as
\begin{equation}
	\hat{\mathbf{H}} = \mathcal{F}\left( \hat{\mathbf{H}}_{\text{LS}} \right) ,
\end{equation}
where $\hat{\mathbf{H}} $ denotes the estimated channel matrix of the neural network.
The DL-based channel inference frameworks are divide into two phases: offline training and online estimation.

\textit{1) Offline Training Phase:}
For training phase, the training-pair set $\mathcal{D}_t $ is fed into the neural network to iteratively update the parameters of $\mathcal{F}$ based on minimizing the loss function.
Then, the training-pair set $\mathcal{D}_t $ are
\begin{equation}
	\mathcal{D}_t = \left\lbrace \left( \mathbf{X}_{\text{in}}^{(1)};\mathbf{X}_{\text{gt}}^{(1)} \right), \cdots, \left( \mathbf{X}_{\text{in}}^{(\|\mathcal{D}_t\|)};\mathbf{X}_{\text{gt}}^{(\|\mathcal{D}_t\|)} \right)  \right\rbrace,
\end{equation}
where $\left( \mathbf{X}_{\text{in}}^{(i)};\mathbf{X}_{\text{gt}}^{(i)} \right) $ denotes the input and ground-truth of the $i$-th training pair and $\|\mathcal{D}_t\|$ is the size of offline training set.

Considering that the data processing of neural network focuses on the real-valued domain, the real part and imaginary part of the complex channel are decomposed and stacked into a 2-feature\footnote{In general, the term ``channel" conventionally refers to data dimensions in feature representations (e.g., RGB channels in images or filter channels in convolutional layers). To avoid confusion with ``channel" of the communication system, we adopt ``feature" as the replacement term throughout this paper.} tensor.
Define the input and corresponding true channel labels as 
\begin{align}
	&\mathbf{X}_{\text{in}}^{(i)}=\left\lbrace \Re(\hat{\mathbf{H}}_{\text{LS}}^{(i)}), \Im(\hat{\mathbf{H}}_{\text{LS}}^{(i)})\right\rbrace \in \mathbb{R}^{N_r\times N_t \times 2} ,\\
	&\mathbf{X}_{\text{gt}}^{(i)}=\left\lbrace \Re({\mathbf{H}}^{(i)}), \Im({\mathbf{H}}^{(i)})\right\rbrace \in \mathbb{R}^{N_r\times N_t \times 2}.	
\end{align}
Next, we employ the empirical mean square error (MSE) criterion during finite offline training time as
\begin{equation}\label{eq:loss}
	\mathop{ \min}\limits_{\mathcal{F}} \frac{1}{\|\mathcal{D}_t\|} \sum_{i=1}^{\|\mathcal{D}_t\|} \left\| \mathbf{X}_{\text{gt}}^{(i)}- \mathcal{F}\left( \mathbf{X}_{\text{in}}^{(i)} \right) \right\|_F^2 .
\end{equation}
Then, it is achieved the well-trained network $\mathcal{F}$.

\textit{2) Online Estimation Phase:}
In this phase, we denote the testing-pair set $\dot{\mathcal{D}}_t$ as
\begin{equation}
	\dot{\mathcal{D}}_t = \left\lbrace \left( \dot{\mathbf{X}}_{\text{in}}^{(1)};\dot{\mathbf{X}}_{\text{gt}}^{(1)} \right), \cdots, \left( \dot{\mathbf{X}}_{\text{in}}^{(\|\dot{\mathcal{D}}_t\|)};\dot{\mathbf{X}}_{\text{gt}}^{(\|\dot{\mathcal{D}}_t\|)} \right)  \right\rbrace,
\end{equation}
where $\left( \dot{\mathbf{X}}_{\text{in}}^{(i)};\dot{\mathbf{X}}_{\text{gt}}^{(i)} \right)$ represents the input and ground-truth of the $i$-th testing pair and $\|\dot{\mathcal{D}}_t\|$ is the size of online testing set.
Then, the testing data $\dot{\mathbf{X}}_{\text{in}}^{(i)}$ sampled from $\dot{\mathcal{D}}_t$ is constructed by the well-trained network.
Denote $\hat{\mathbf{H}}_{\text{net}}^{(i)}$ as the estimated channel corresponding to the $i$-th sample, the estimated channel tensor $\dot{\mathbf{X}}_{\text{est}}$ can be given by
\begin{equation}
	\dot{\mathbf{X}}_{\text{est}}^{(i)} = \mathcal{F}_{\text{net}}\left( \dot{\mathbf{X}}_{\text{in}}^{(i)} \right),
\end{equation}
where $\dot{\mathbf{X}}_{\text{est}} = \left\lbrace \Re(\hat{\mathbf{H}}_{\text{net}}^{(i)}), \Im(\hat{\mathbf{H}}_{\text{net}}^{(i)}) \right\rbrace  \in \mathbb{R}^{N_r \times N_t \times 2}$.
Note that the testing data is independent on the offline training data, which is employed for exclusive evaluation of the performance of the well-training network.

\textit{Remark 2:}
According to the exploration on antenna correlations $\mathbf{R}_\theta$ and $\mathbf{R}_r$ in Section \ref{section:antenna correlation}, the non-linearity of the near-field channel phase causes the number of strongly correlated elements surrounding each channel element to vary with its position.
While CNN is effective in processing local contexts, its fixed receptive fields constrain its ability to capture the intricate and non-uniform spatial features of near-field channels. 
In contrast, transformer excels at in-context learning (ICL), leveraging the self-attention mechanism to capture relationships across all elements in the input sequence, regardless of their distance \cite{Guo2023_How}.
This capability is particularly advantageous for ELAA to capture interactions between the elements of short range or long range.

Antenna elements exhibit strong correlation with adjacent elements, as indicated by $\mathbf{R}_\theta$ and $\mathbf{R}_r$. 
Therefore, the channel is divided into several subchannels and learning the correlations of inter-subchannel instead of inter-element.
We develop a multi-scale spatial attention network (MsSAN) to concentrate on correlations between subchannels rather than all antenna elements to generate effectively context-aware representations of the near-field channel matrix.


\begin{figure*}[!htbp]
	\centering
	\includegraphics[width=0.90\textwidth]{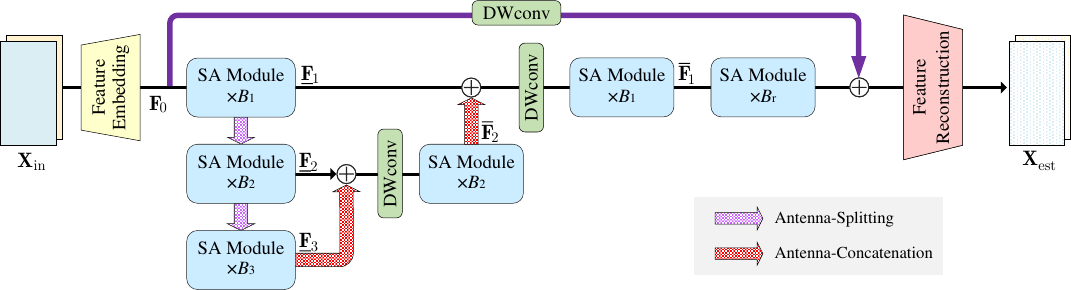}
	\caption{The three-layer architecture of multi-scale spatial attention network for channel estimation (MsSAN).}
	\label{fig:mssan}
\end{figure*}

\begin{figure*}[htbp]
	\centering
	\includegraphics[width=0.62\textwidth]{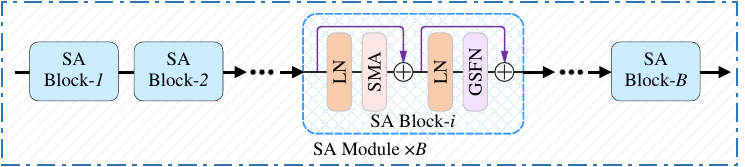}
	\caption{The spatial attention module consists of $B$ consecutive SA blocks.
	}
	\label{fig:samodule}
\end{figure*}

\subsection{The Structure Design of MsSAN}

Following the antenna correlations, we employ the multi-scale architecture to obtain subchannels of different size and design the spatial attention (SA) module to capture the interactions of subchannels, presented in Fig. \ref{fig:mssan}.
Our network employs a three-layer architecture implemented by antenna-splitting and antenna-concatenation operations to achieve multi-scale information extraction.
The attention module of each layer focuses on subchannels of different sizes.
Namely, the attention is paid to form global channel features of higher layers to more fine-grained subchannel features of lower layers.
Then, we first introduce the detailed structure of the proposed network.

Given a noisy channel tensor $\mathbf{X}_{\text{in}} \in \mathbb{R}^{N_r \times N_t \times 2} $, the embedding module is first utilized to extract the $C$-feature primitive features $\mathbf{F}_0$ by a feature embedding operation $f_{EB}(\cdot)$ with a $\{3,1,C\}$\footnote{For the sake of conciseness in the paper, denote $\{a, b, c\}$ as the convolutional layer, where $a$ is a $a\times a$ filter, $b$ is a stride of $b$ and $c$ is the number of out features.}.
Thus, the dimension of the input is converted from $\mathbb{R}^{N_r  \times N_t \times 2}$ to $\mathbb{R}^{N_r  \times N_t \times C}$, i.e., $\mathbf{F}_0 \in \mathbb{R}^{N_r  \times N_t \times C} $.
Then, the embedding tensor $\mathbf{F}_0$ travels 3-layer symmetric encoder-decoder, where different layers focus on different scales features.
In detail, the $i$-th layer encoder/decoder is implemented by a spatial attention (SA) module.
As shown in Fig. \ref{fig:samodule}, the SA module of the $i$-th layer consists of $B_i$ consecutive spatial attention (SA) blocks, defined as $f_{SA}^{(B_i)}(\cdot)$ for $i\in \{1, 2, 3\}$.
Furthermore, SA modules are designed to focus on extracting the correlation between near-field subchannels, as detailed in Section \ref{subsection:sa module}.

The SA modules between the higher layers and lower layers are cascaded through antenna-splitting operations $f_{AS}(\cdot)$ in the encoder, enabling each layer to focus on distinct regions of the features.
The antenna-splitting operation halves the receiving antenna dimension of the tensors while doubling the number of features, whose implementation is introduced in detail in next.
Overall, the encoder procedure can be formulated as
\begin{align}\label{eq:encoder}
	&\underline{\mathbf{F}}_1 = f_{SA}^{(B_1)}(\mathbf{F}_0),
	\\
	&\underline{\mathbf{F}}_2 = f_{SA}^{(B_2)} \left( f_{AS}(\underline{\mathbf{F}}_1)\right),
	\\
	&\underline{\mathbf{F}}_3 = f_{SA}^{(B_3)} \left( f_{AS}(\underline{\mathbf{F}}_2)\right),
\end{align} 
where $\underline{\mathbf{F}}_1 \in \mathbb{R}^{N_r \times N_t \times C}$ and $\underline{\mathbf{F}}_2 \in \mathbb{R}^{\frac{N_r}{2} \times N_t \times 2C}$  are the output tensors of SA blocks in the first and second layers, respectively.
$ \underline{\mathbf{F}}_3 \in \mathbb{R}^{\frac{N_r}{4} \times N_t \times 4C} $ is the output of encoder.

With the purpose of recovery, antenna-concatenation and `shortcut connection' are employed to enable cross-layer feature fusion with the previous layer.
Furthermore, these preserve the integrity of the information and facilitate better capture of the finer details within the channel data.
Contrary to the antenna-splitting, the antenna-concatenation operation $f_{AC}(\cdot)$ doubles the receiving antenna dimension of the tensor while halving the number of features, whose implementation is introduced in detail in next.
Therefore, the decoder process is
\begin{align}
	&\overline{\mathbf{F}}_2=f_{SA}^{(B_2)} \left( f_{DW} \left(\underline{\mathbf{F}}_2 + f_{AC}(\underline{\mathbf{F}}_3) \right)\right),\\
	&\overline{\mathbf{F}}_1=f_{SA}^{(B_1)} \left( f_{DW} \left(\underline{\mathbf{F}}_1 + f_{AC}(\overline{\mathbf{F}}_2) \right)\right),
\end{align}
where $\overline{\mathbf{F}}_1 \in \mathbb{R}^{N_r  \times N_t \times C}$ and $\overline{\mathbf{F}}_2 \in \mathbb{R}^{\frac{N_r}{2} \times N_t \times 2C}$  are the output tensors of SA blocks of first and second layers, respectively.
The DWconv $f_{DW}(\cdot)$ is employed to refine and fuse the information of cross-layers.
Next, $B_r$ SA blocks are employed in the tail to refine and enrich the information of $\overline{\mathbf{F}}_1$.

In a feed-forward network (FFN), we adopt the residual structure by `skip connection'. Additionally, a single depth-wise convolution (DWconv) layer with a $3 \times 3$ filter
is employed in the skip connection for emphasizing on the local context of the feature map before arriving the feature reconstruction module in the tail.
The feature reconstruction module in the tail is employed to reduce tensors back to the original input dimension, realized by a convolutional layer $f_{RC}(\cdot)$ by $\{3,1,2\}$.

In total, given input channel tensor $\mathbf{X}_{\text{in}}$, the final output of MsSAN $\mathcal{F}_{MsSAN}(\mathbf{X}_{\text{in}})$ can be expressed as
\begin{equation}
	\mathbf{X}_{\text{est}} = f_{RC}\left( f_{DW}(\mathbf{F}_0)+ f_{SA}^{(B_r)}(\overline{\mathbf{F}}_1) \right) ,
\end{equation}
where $\mathbf{X}_{\text{est}} \in \mathbb{R}^{N_r \times N_t \times 2}$ is the estimated channel tensor of the entire network.

\subsection{Antenna-Splitting and Antenna-Concatenation}
\begin{figure}[t]
	\centering
	\begin{subfigure}[b]{0.49\textwidth}
		\centering
		\includegraphics[width=\textwidth]{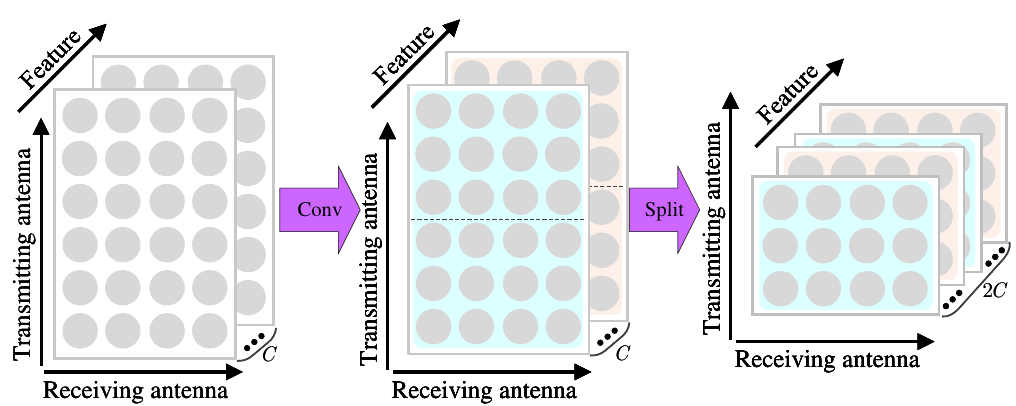}
		\caption{Antenna-Splitting}
		\label{fig:as}
	\end{subfigure}
	\hfil
	\begin{subfigure}[b]{0.49\textwidth}
		\centering
		\includegraphics[width=\textwidth]{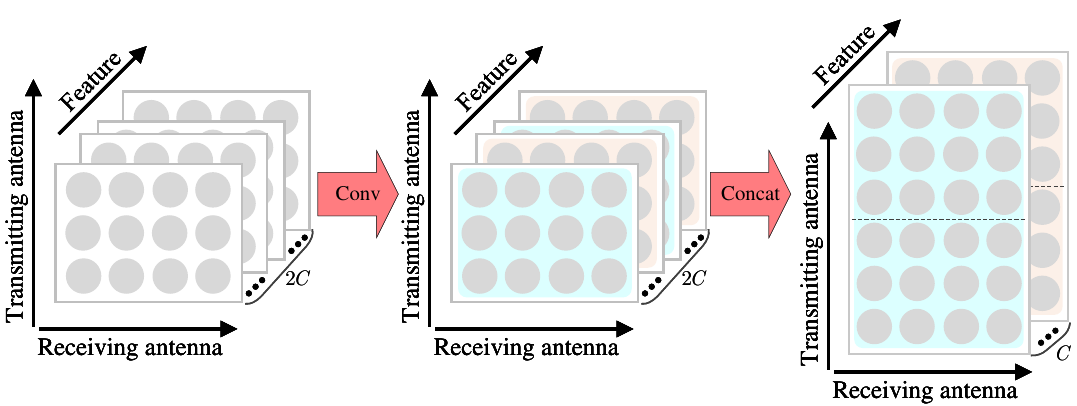}
		\caption{Antenna-Concatenation}
		\label{fig:ac}
	\end{subfigure}
	\caption{Illustration of information flow of the antenna-splitting from layer $1$ to layer $2$ and the antenna-concatenation from layer $2$ to layer $1$.
	The gray circles represent the elements of the near-field channel features.}
	\label{fig:sampling}
\end{figure}
The antenna-splitting and antenna-concatenation are designed to refine the subchannels of near-field channel targeted by SA modules.

The antenna-splitting is used to arrange the subchannel features of the near-field channel into the feature dimension.
In detail, we utilize a convolutional layer with a $3\times 3$ filter and a splitting operation.
The convolutional layer is used to fuse and extract feature information with remaining the feature number. 
The splitting operation distributes the half of channel features from the receiving antenna dimension to the feature dimension.
The information flow of the antenna-splitting from layer 1 to layer 2 is illustrated as in Fig. \ref{fig:as}.

The antenna-concatenation is used to arrange the features into the near-field channel spatial dimension.
In detail, the antenna-concatenation operation utilizes a $3\times 3$ convolutional layer and concatenates the half of feature dimension into the receiving antenna dimensions.
The information flow of the antenna-concatenation from layer 2 to layer 1 is illustrated as in Fig. \ref{fig:ac}.

In the first layer, a whole channel feature is viewed as a subchannel feature.
In the second layer, the half of channel feature is viewed as a subchannel feature.
In the third layer, the quarter of channel feature is viewed as a subchannel feature. 
Overall, antenna-splitting refines channel features from the broad global subchannel to more fine-grained subchannels and vice versa for antenna-concatenation.
Therefore, the first layer attends to the global correlation of the spatial antennas, while the third layer concentrates on the finer details of smaller antenna regions.


\begin{figure}[t]
	\centering
	\includegraphics[width=0.45\textwidth]{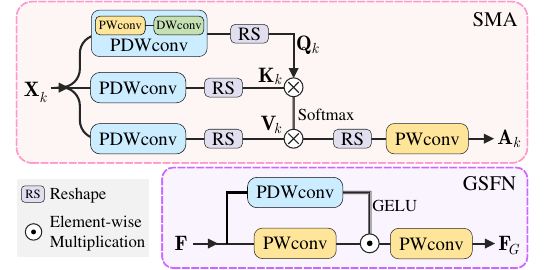}
	\caption{The details of the spatial multi-head attention (SMA) and the gated spatial feed-forward network (GSFN).}
	\label{fig:net_detail}
\end{figure}
\subsection{The Implementation of the Spatial Attention Module}\label{subsection:sa module}
The spatial attention (SA) module focuses on interactions between subchannel features.
As shown in Fig. \ref{fig:samodule}, each module with two skip connections consists of  three components: layer normalization (LN) layer, the spatial multi-head attention (SMA) and a gated spatial feed-forward network (GSFN).
Note that we employ layer normalization (LN) layer instead of batch normalization layer, which aims to preserve instance details and accelerate convergence.
Next, SMA and GFFN are introduced with details in the following.

\textit{1) Spatial Multi-head Attention (SMA)}

Given that antenna elements always maintain strong correlation with neighboring antenna elements, we concentrate on correlation of inter-subchannel.
This avoids the high computational complexity caused by calculating the correlation between all the antennas.
Hence, we design the spatial multi-head attention (SMA) tailored to assign importance by computing attention weights between subchannels, where the sum of dot products is introduced as SA map of subchannels rather than the cross-covariance across features of the general attention.

In this work, we apply the multi-head attention mechanism to explore the dependencies and information among the features and divide the features into the number of heads $K_i$, where $K_i$ is the number of heads in the $i$-th layer for $i\in \{1,2,3\}$.

For the sake of notation, omit the subscript $i$.
Let the tensor from the first LN layer be $\mathbf{X}\in \mathbb{R}^{\tilde{H}\times \tilde{W}\times\tilde{C}}$.
Then, $\mathbf{X}$ is divided into $K$ tensors $\{\mathbf{X}_k \in \mathbb{R}^{\tilde{H}\times \tilde{W}\times\tilde{C}_k}  \}_{k=1}^{K}$.
Subscript $k\in \{1,\cdots,K\}$ represents the head of the attention and $\sum_{k} \tilde{C}_k=\tilde{C}$.
The $1$-head attention mechanism is presented in Fig. \ref{fig:net_detail}.

\begin{figure}[t]
	\centering
	\includegraphics[width=0.35\textwidth]{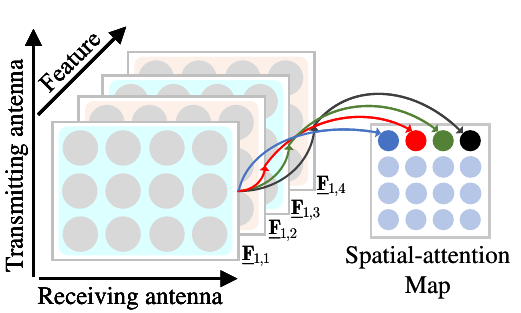}
	\caption{Illustration of the information transmission of the attention mechanism.
	The features with the same color are different sunchannel features of a channel feature and pieced together into a whole channel feature. }
	\label{fig:attention}
\end{figure}

After the tensor $\mathbf{X}_k \in \mathbb{R}^{\tilde{H}\times \tilde{W}\times\tilde{C}_k}$ arrives firstly point-depth-wise convolutions (PDWconv) $f_{PDW}(\cdot) $, the tensor is reshaped to $\tilde{H}\tilde{W}\times \tilde{C}_k$ by reshaping operations $RS(\cdot)$.
Then, we have query subspace $\mathbf{Q}_k \in \mathbb{R}^{\tilde{H}\tilde{W}\times \tilde{C}_k}$, key subspace $\mathbf{K}_k\in \mathbb{R}^{\tilde{H}\tilde{W}\times \tilde{C}_k}$ and value subspace $\mathbf{V}_k\in \mathbb{R}^{\tilde{H}\tilde{W}\times \tilde{C}_k}$ for the $k$-th head.
A PDWconv is composed of cascaded a point-wise convolution (PWconv) by $\{1,1,\tilde{C}_k\}$ and a DWconv and emphasizes on the local context of features.
The reshaping operations $ \underline{RS} (\cdot)$ is utilized to reshape the output of convolutions to $\tilde{H}\tilde{W}\times \tilde{C}_k $ to operate sum of dot products.

The spatial-attention map $\mathbf{M}_k\in \mathbb{R}^{\tilde{C}_k\times \tilde{C}_k} $ can be calculated by sum of dot products and written as $\mathbf{M}_k = \mathbf{K}_k^T \mathbf{Q}_k$.
Then, the SMA process is defined as
\begin{equation}
	\begin{aligned}
		&\mathbf{A}_k = f_{PW}(\overline{RS}(\mathbf{V}_k \cdot \mathrm{Softmax}({\mathbf{M}_k}/{W_k}))),\\
		&\mathbf{A}=\mathrm{concat}(\{\mathbf{A}_k\}_{k=1}^K),
	\end{aligned}
\end{equation}
where $\mathbf{A}_k \in \mathbb{R}^{\tilde{H}\times \tilde{W}\times\tilde{C}_k}$ is output of the $k$-th head, $W_k$ is a learnable scaling parameter to prevent saturation the following softmax function and $\overline{RS}(\cdot)$ is used to reshape the tensor to size $\tilde{H}\times \tilde{W}\times\tilde{C}_k $. 
Then, all heads of the multi-head attention $\{\mathbf{A}_k\}_{k=1}^K$ are concatenated to a tensor $\mathbf{A}\in \mathbb{R}^{\tilde{H}\times \tilde{W}\times\tilde{C}}$.
In the tail of SMA, PWconv promotes information across different features and capture more complex feature combinations.

In essential, the attention map $\mathbf{M}$ is interpreted as a correlation weight matrix of the subchannel features.
To illustrate this, we take the second layer of the encoder as an example, $\mathbf{X}$, as shown in Fig. \ref{fig:attention}, assume the embedding feature $C$ is $2$.
After traveling the antenna splitting, the tensor is expressed as $\{\underline{\mathbf{F}}_{1,1}, \underline{\mathbf{F}}_{1,2}, \underline{\mathbf{F}}_{1,3}, \underline{\mathbf{F}}_{1,4}\} \in \mathbb{R}^{\frac{N_r}{2}  \times N_t \times 4}$, where $\underline{\mathbf{F}}_{1,c} \in \mathbb{R}^{\frac{N_r}{2}  \times N_t}$ for $c\in\{1,2,3,4\}$.
Actual, $\{\underline{\mathbf{F}}_{1,1}, \underline{\mathbf{F}}_{1,2} \} $  and $\{\underline{\mathbf{F}}_{1,3}, \underline{\mathbf{F}}_{1,4} \} $ are the upper subchannel and lower subchannel feature of tensor $\underline{\mathbf{F}}_1 $.
Then, the attention module computes the correlation weight across these subchannel features.
The $(i,j)$-th entry of spatial attention map $\mathbf{M}$ can be interpreted as attention weight of $\mathrm{vec}(\underline{\mathbf{F}}_{1,i})^T \mathrm{vec}(\underline{\mathbf{F}}_{1,j})$ by calculated with SMA.
$\mathrm{vec}(\underline{\mathbf{F}}_{1,i})^T \mathrm{vec}(\underline{\mathbf{F}}_{1,j})$ for $i\in \{1,3\}$ and $j\in \{2,4\}$ is represented as the correlation between lower and upper subchannel features.
$\mathrm{vec}(\underline{\mathbf{F}}_{1,i})^T \mathrm{vec}(\underline{\mathbf{F}}_{1,j})$ for $i\in \{1,3\}$ and $j\in \{1,3\}$ is represented as the autocorrelation between lower subchannel features.

The focused subchannels are refined layer by layer through the antenna-splitting.
Therefore, the SMA pays attention on correlations between subchannel features to adjust the network to recovery channel accurately.

\textit{2) Gated Spatial Feed-Forward Network (GSFN)}

In transformer networks, a feed-forward network (FFN) is an essential part of enhancing feature representation.
We integrate simple attention and gated linear unit (GLU) into the proposed GSFN to capture spatial information more efficiently.
We adopt PWconv and PDWconv layers to generate feature map, as shown in Fig. \ref{fig:net_detail}. 

Given the tensor $\mathbf{F} \in  \mathbb{R}^{\tilde{H}\times \tilde{W}\times\tilde{C}}$ from the second LN layer, the process of GSFN can be represented as
\begin{equation}
	\mathbf{F}_G = f_{PW}(\mathrm{GELU}(f_{PDW}(\mathbf{F})) \odot f_{PW}(\mathbf{F})),
\end{equation}
where $f_{PDW}(\cdot) $ is the convolution cascaded with PWconv by $\{1,1, \tilde{C} \}$ and DWconv by $\{3,1, \tilde{C}\}$.
$\mathrm{GELU}(\cdot)$ is a GELU activation function.
$f_{PW}(\cdot)$ is a PWconv by $\{1,1,\tilde{C}\}$ to adjust the features to the input dimension.

In GSFN, DWconv emphasizes on the local context of neighboring entries for learning communication channel information efficiently.
Especially, this unit focuses the attention on the continuity of features with the element-wise multiplication, which complements the SMA's lack of attention to  individual elements.

%
%

\section{Numerical Results}\label{section:simulation}
In this section, we present evaluations of the proposed transformer-based schemes for near-field channel estimation.

\begin{table}[t] 
	\begin{center} 
		\caption{The Simulation Dataset Configurations}
		\label{table:datasetConfig}
		\begin{tabular}{c c}
			\toprule[0.5mm]
			\textbf{Parameters} & \textbf{Value}\\
			\toprule[0.25mm]
			The number of  antennas $N_r, N_t$ & $256, 8$ \\
			Carrier frequency $f_c$ & $60$ GHz \\
			The number of paths $L$ & $L\sim Poisson(6)$\\
			The distribution of angle $\phi$ & $\phi \sim \mathcal{U}(-\frac{\pi}{3}, \frac{\pi}{3}) $\\
			The Rayleigh distance $d_R$ & $174.25$ m \\
			The distance between Rx and Tx  $r$ & $r\sim \mathcal{U}(3, d_R)$ m\\
			Channel model realization & Extended Saleh-Valenzuela\\
			Signal-to-noise ratio SNR（(dB)& $\{-10,-5,0,5,10\}$\\
			Dataset split (train:test) & $4:1$\\
			\toprule[0.5mm]
		\end{tabular}
	\end{center}
\end{table}

\begin{table}[t] 
	\begin{center} 
		\caption{Settings of the Neural Networks}
		\label{table:networkParam}
		\begin{tabular}{c c}
			\toprule[0.5mm]
			\textbf{Parameters} & \textbf{Value}\\
			\toprule[0.25mm]
			Embedding features $C$ & $32$\\
			SA blocks $\{B_1, B_2, B_3, B_r\}$ & $\{1,1,2,1\}$ \\
			Heads $\{K_1,K_2, K_3, K_r\}$ & $\{1,2,4,1\}$\\
			Training epoch & $120$\\
			Batch size & $32$\\
			Learning rate $\gamma_t$ & $\gamma_0=0.1$ with warmup\\
			Scheduler & Cosine annealing strategy\\
			Training optimizer& SGD with momentum $0.9$\\
			Weight decay& $1\times 10^{-4}$\\
			\toprule[0.5mm]
		\end{tabular}
	\end{center}
\end{table}

\begin{figure}[t]
	\centering
	\includegraphics[width=0.42\textwidth]{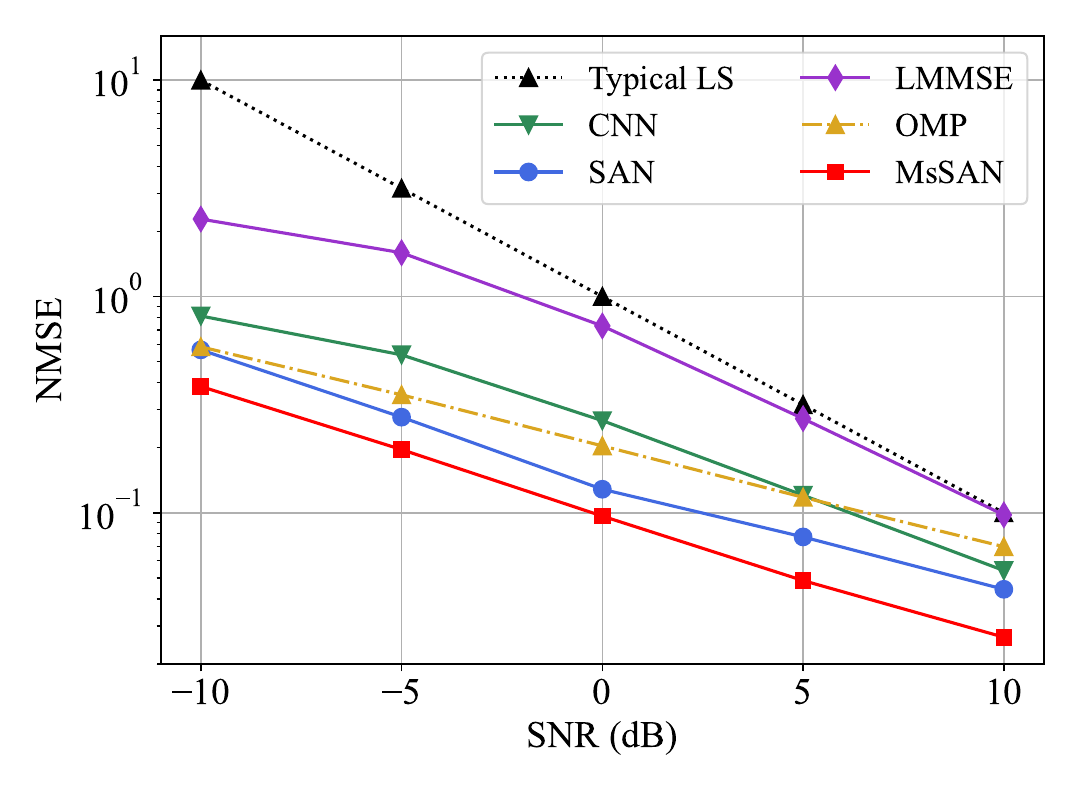}
	\caption{
		Channel estimation performance of our schemes compared to others in different SNR regimes.}
	\label{fig:snr_nmse}
\end{figure}

\subsection{Simulation Setups}
The system default configurations are shown as follows:
The receiver and transmitter are equipped with $N_r=256$ and $N_t = 8$, respectively.
The carrier frequency is $f_c=60$ GHz.
The distances between the transmitter and receiver are uniformly chosen from $[3, d_R]$, i.e., $r\sim \mathcal{U}(3, d_R)$.
The paths number of the near-field channels follows Poisson distribution with mean $6$, i.e., $L\sim Poisson(6)$.
The physical AoA/AoD of each path are chosen randomly from $[-\frac{\pi}{3}, \frac{\pi}{3}] $, i.e., $\phi \sim \mathcal{U}(-\frac{\pi}{3}, \frac{\pi}{3})$。
Additionally, the signal-to-noise ratio (SNR) is defined as $\text{SNR} = 1/\sigma^2$.
Then, the channel dataset is generated by employing extended Saleh-Valenzuela model based on the detailed environment parameters summarized in Table \ref{table:datasetConfig}.
The size of samples in the training set and testing set are split by $4:1$.
It is noticed that the testing set is independent on the training set in the offline training phase and is used to evaluate the performance of the proposed scheme.
Furthermore, the detailed neural network training settings are all based on Table \ref{table:networkParam}.
As for the neural network training, we utilize the SGD optimizer with momentum $0.9$ and weight decay $1\times 10^{-4}$ and set the batch size $32$.
Then, the training learning rate are scheduled by
\begin{equation}
	\gamma_t= 
	\begin{cases}
		\dfrac{\gamma_0}{6-t}, & 1 \leq t \leq 5, \\ 
		\dfrac{\gamma_0}{2}\left(1+\cos \left(\dfrac{(t-5) \pi}{T-4}\right)\right), & 6 \leq t \leq T,
	\end{cases}
\end{equation}
The learning rate is set increase linearly in the first 5 epochs and update based on cosine annealing strategy.

The normalized mean-squared error (NMSE) is selected as the performance metric, which is defined as
\begin{equation}
	\mathrm{NMSE} = \mathbb{E}\left\lbrace 
	  \frac{\lVert \mathbf{H}- \hat{\mathbf{H}}  \rVert_{F}^{2} }{\lVert \mathbf{H} \rVert_{F}^{2}} 
	\right\rbrace ,
\end{equation}
where $\mathbf{H}$ and $\hat{\mathbf{H}}$ are the true channel matrix and the estimated channel matrix, respectively.
Besides, we also consider the spectral efficiency (SE) as the other metric, which is widely used to evaluate the estimation performance \cite{Zhang2024_HDnGAN}.
Tx and Rx are treated as the base station and user equipment, respectively.
In TDD wireless system, due to the channel reciprocity, the estimated channel is adopted to implement the downlink transmission involving the MRT scheme.
The spectral efficiency is defined as 
\begin{equation}
	\mathrm{SE} = \log_2[ 1+ \frac{\mathrm{tr}(\hat{\mathbf{H}} \mathbf{H}^H \mathbf{H} \hat{\mathbf{H}}^H)}{\sigma^2 \mathrm{tr}(\hat{\mathbf{H}} \hat{\mathbf{H}}^H )} ]. 
\end{equation}
Then, we evaluate performance of the proposed channel inference algorithms compared to other algorithms:
\begin{itemize}
	\item LS and Linear MMSE (LMMSE);
	\item Orthogonal matching pursuit (OMP)-based XL-MIMO channel estimation \cite{Lu2023_Field,NearCE_Dai}, termed as OMP;
	\item Deep CNN-based channel estimation \cite{Xu2024_Deep,Lee2022_Intelligent}, termed as CNN;
	\item The proposed multi-scale spatial attention network for channel estimation, termed as MsSAN.
	\item The single-scale spatial attention network is regraded as the ablation experiment for our scheme, termed as SAN. Notice that SAN employs $10$ SA blocks and $C=48$ the embedding features in total.
\end{itemize}

\subsection{Performance Comparison with Existing Methods}

\begin{figure}[t]
	\centering
	\includegraphics[width=0.42\textwidth]{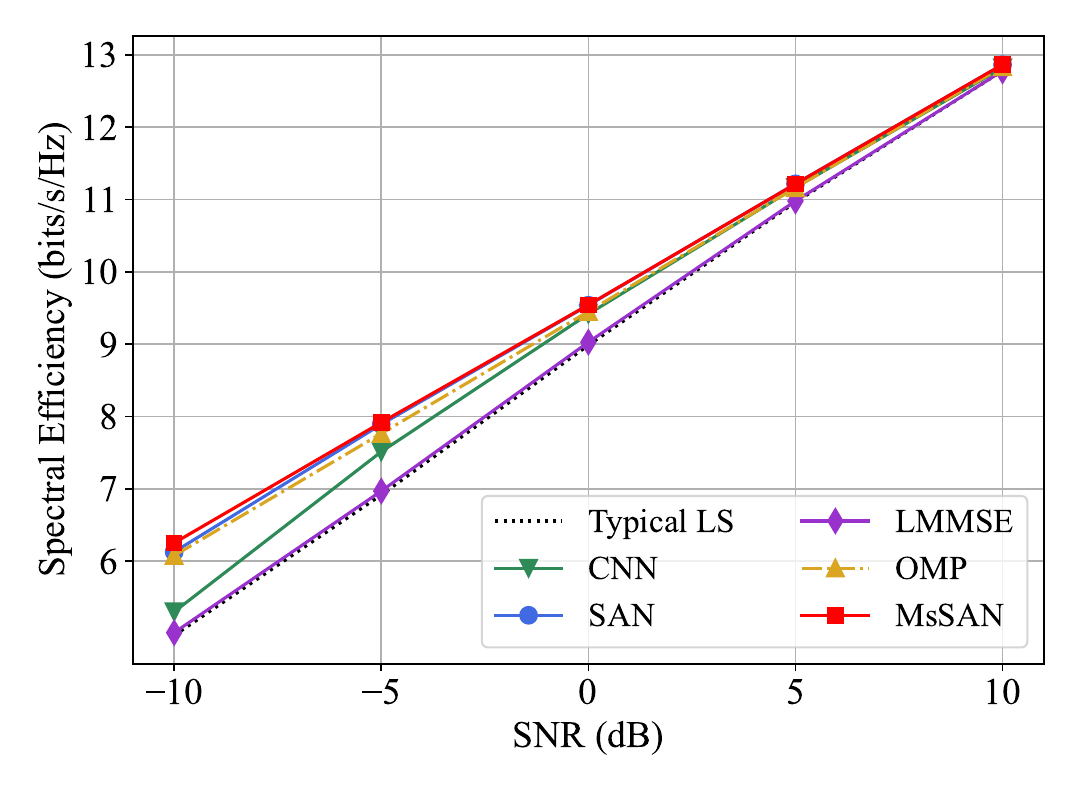}
	\caption{
		The spectral efficiency performance of different schemes across SNRs.}
	\label{fig:snr_se}
\end{figure}

\begin{figure}[t]
	\centering	
	\includegraphics[width=0.42\textwidth]{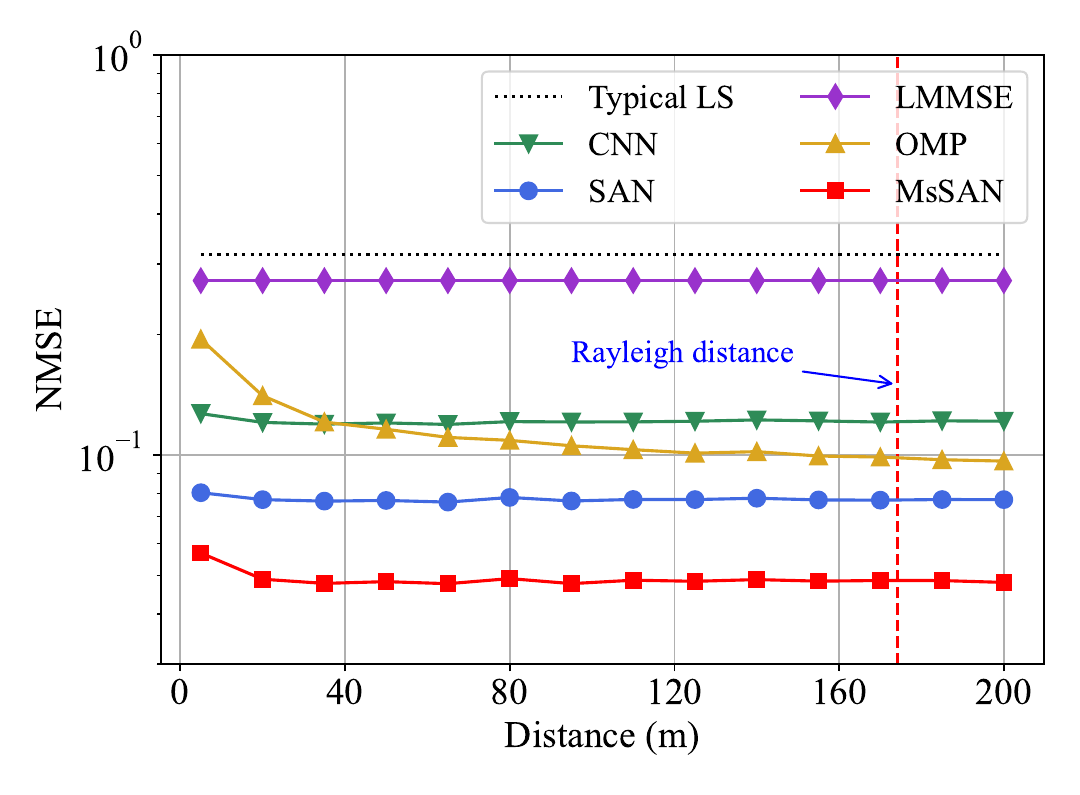}
	\caption{Comparison of the NMSE performance of different schemes in different distance with SNR= $5$ dB and $L\sim Poisson(6)$.} 
	\label{fig:dis_nmse}
\end{figure}


\begin{figure*}[t]
	\centering
	\includegraphics[width=0.70\textwidth]{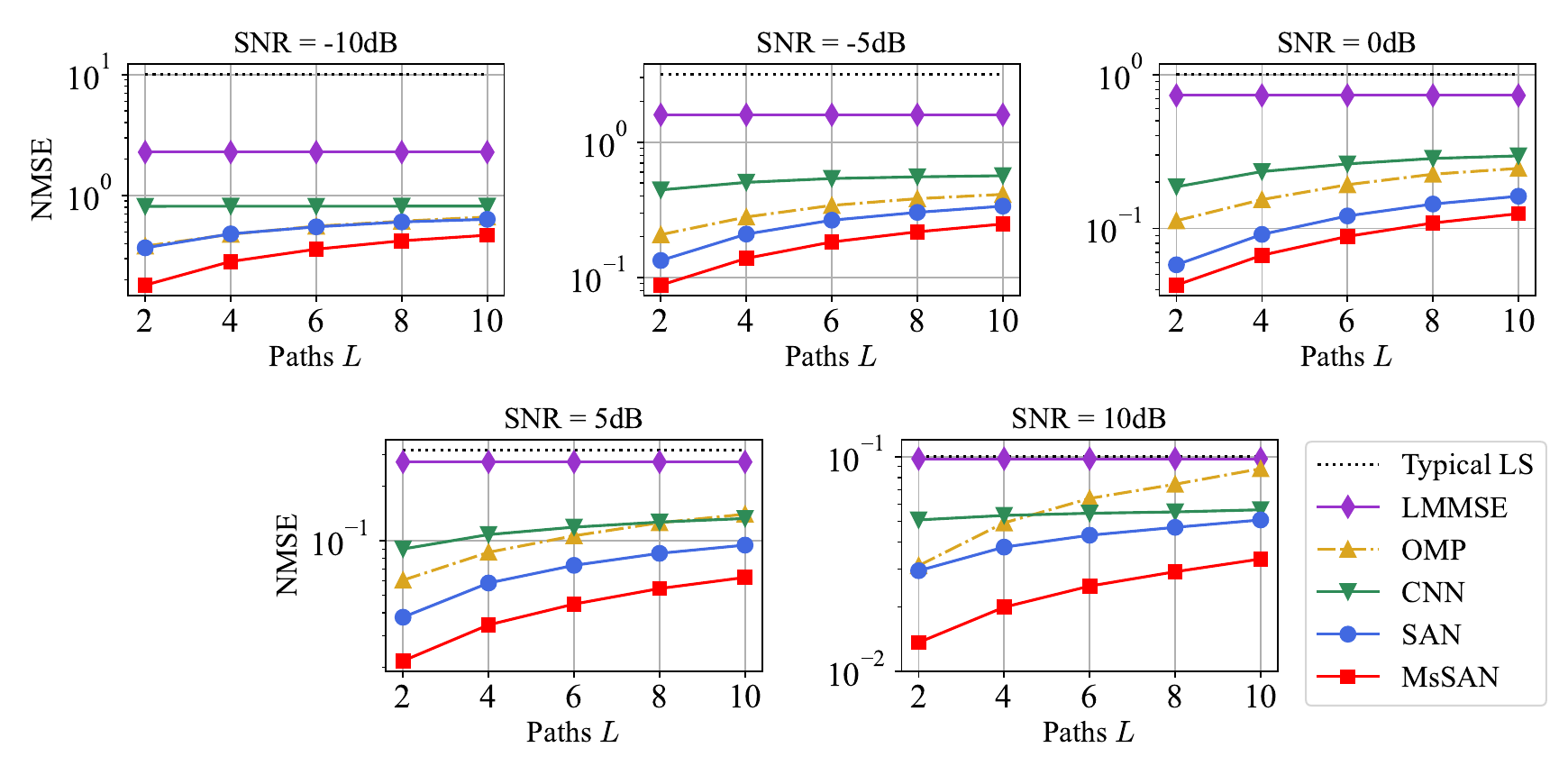}
	\caption{
		The NMSE performance against the number of paths at different SNRs when different channel estimation schemes are used.}
	\label{fig:path_nmse}
\end{figure*}
Here, we evaluate the performance of our proposed method under different SNR scenarios, as shown in Fig. \ref{fig:snr_nmse}.
It is observed that the schemes based on transformer structure significantly outperform other counterparts, especially our proposed MsSAN exhibits the best NMSE performance.
As for LS estimator, it suffers from the poor performance, particularly in the low-SNR regime. 
Although LMMSE estimator performs better than LS in the low-SNR regime, it is still unsatisfactory compared to other estimation schemes.
This is because both LS and LMMSE estimators rely on a linear formulation of channel reconstruction, without exploiting the inherent characteristics of XL-MIMO channels.
It is worth noting that the CNN-based estimator performs worse than the OMP-based scheme, particularly in the low-SNR regime.
As SNR increases, CNN gradually surpasses OMP by virtue of its superior image processing capabilities.
The OMP scheme is a sparse signal recovery method, whose exceptional performance is attributed to a carefully designed dictionary tailored for near-field channels.
CNN treats the channel matrix as an image for processing and denoising, disregarding the inherent properties of the input.
Due to its fixed receptive field, CNN struggles to accurately capture the spatial features across channel elements.
In contrary, both SAN and MsSAN leverage the self-attention mechanism to adaptively focus the intricate and non-uniform spatial features across channel features.
Meanwhile, the NMSE performance gaps between OMP and SAN are very narrow in the low SNR regime and MsSAN significantly outperforms SAN and OMP.
The single-scale architecture weakens the correlation between adjacent subchannels captured by SAN.
Multi-scale spatial attention by segmenting channel features layer by layer allow MsSAN to compute the spatial correlation between subchannel features such that MsSAN capture both fine-grained local features and broader global features  to more accurate channel estimation.

The spectral efficiency across different SNRs is presented when different channel estimation schemes are implemented, as depicted in Fig. \ref{fig:snr_se}.
It is clear that the SE performance of our proposed near-field channel denosing method is superior to others.
Due to the weakness of CNN in the low-SNR regime, its performance surpasses only LS and LMMSE.
This phenomenon further illustrates the non-negligible limitations of CNN being directly applied to channel estimation in the absence of considering near-field channel characteristics.
As for OMP scheme, its SE performance is relatively close to that of SAN and MsSAN, but it is highly dependent on the selection of the near-field dictionary.
This disadvantage becomes more apparent in next simulations against distance and paths.
The inter-subchannel correlation learning capabilities enables MsSAN to achieve better SE performance than SAN.



To validate the robustness of our proposed networks, we conduct simulations with various distances.
In our simulation, the Rayleigh distance $d_R$ equals $174.25$ meters and SNR is fixed as $5$ dB.
The distance range between Tx and Rx varies from $5 $ meters to $200$ meters.
Notice that the evaluated neural networks are trained only on data within the near-field region during offline training phase.
The NMSE performance against different distances is depicted in Fig. \ref{fig:dis_nmse}. 
It is evident that our proposed method consistently outperform other approaches across all distances while maintaining better performance in both the near-field and far-field regions.
In addition, when the Rx and Tx are in very close proximity, all schemes except LS and LMMSE exhibit some degree of performance degradation, with the OMP scheme showing the most noticeable decline. 
This issue of OMP scheme stems from the imperfect design of the dictionary, as dictionary vectors fail to achieve perfect orthogonality in the distance domain.
Hence, near-field dictionary design has been a key bottleneck in near-field estimation.
Thanks to the attention mechanisms in SMA and GSFN, both SAN and MsSAN maintain exceptional performance even when the communication distance is very close.
The combination of the SA and multi-layer design makes MsSAN exhibit higher robustness and generalization across different distances, even in the far-field range.

To evaluate the stability of our estimation schemes in different communication environments, we conduct experimental verification under paths $L\in\{2,4,6,8,10\}$ at different SNRs.
As illustrated in Fig. \ref{fig:path_nmse}, it is evident that fewer paths in the channel make the estimation schemes perform better, except for LS and LMMSE.
It is because simpler channels have fewer parameters and less complex structures, making estimators more straightforward to reconstruct channels accurately.
When SNR is -10 dB, the NMSE of SAN and OMP overlap, which shows the weakness of SAN in low SNR regimes.
Clearly, our proposed MsSAN shows the most outstanding NMSE performance across different paths and varying SNR conditions.
Overall, it showcases robustness and adaptability of our schemes in near-field communication scenarios.

\begin{figure*}[t]
	\centering
	\includegraphics[width=0.7\textwidth]{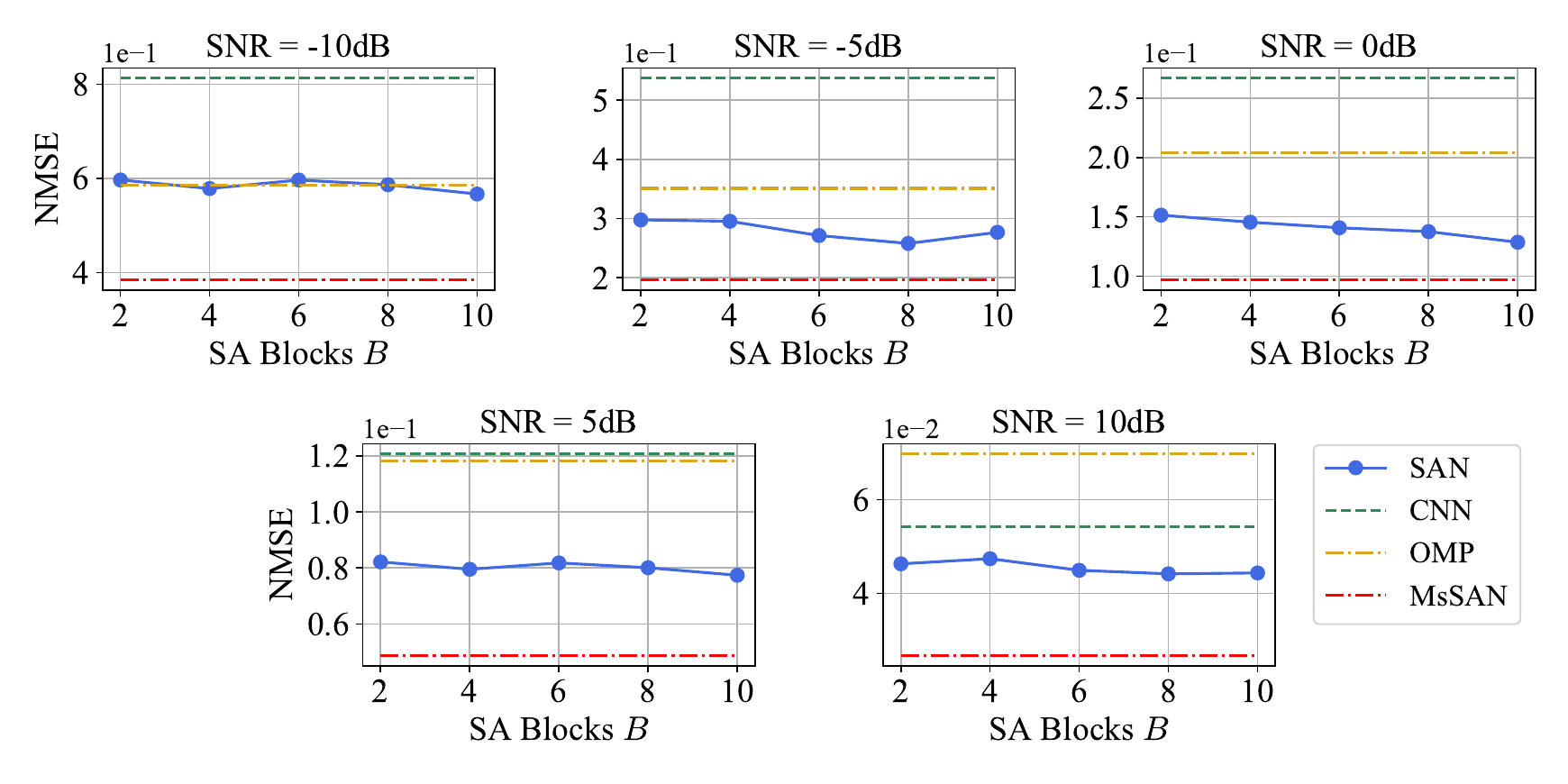}
	\caption{
		The NMSE performance of SAN against SA blocks at different SNRs, where the number of SA blocks is fixed.}
	\label{fig:block_nmse}
\end{figure*}

\begin{figure}[t]
	\centering
	\includegraphics[width=0.42\textwidth]{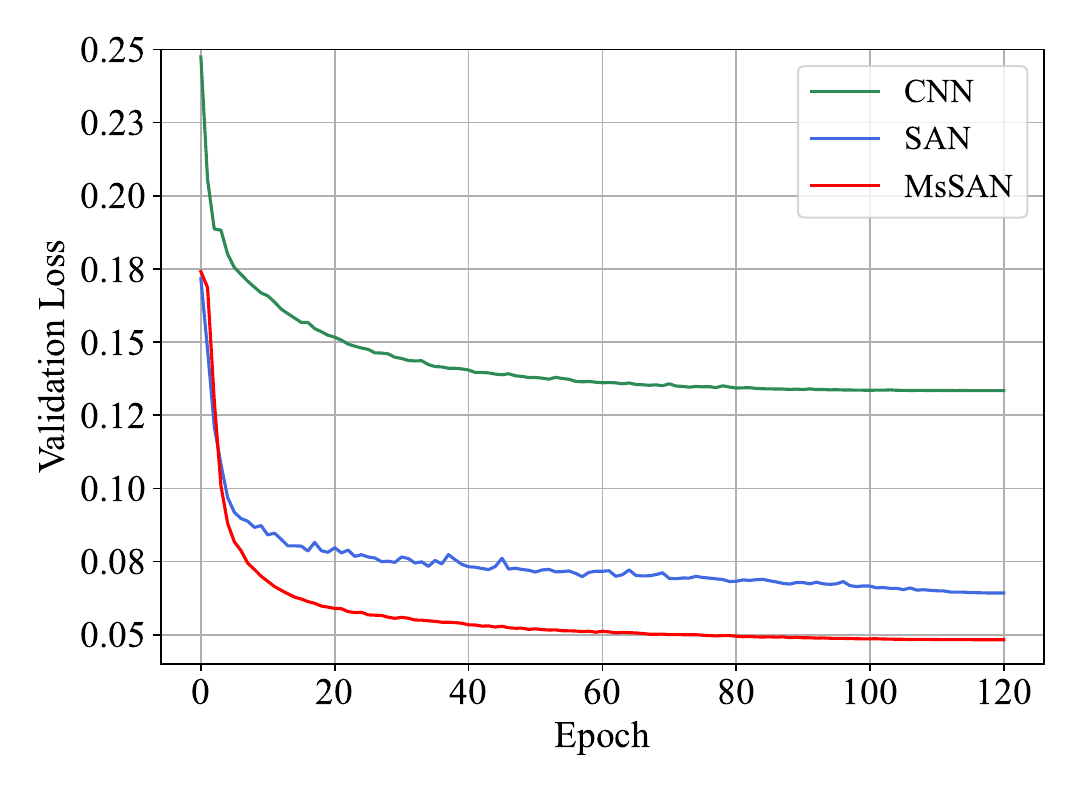}
	\caption{
		The convergence performance comparison of different network structures.}
	\label{fig:loss}
\end{figure}

\subsection{Performance with Different Number of Blocks and Loss Convergence}
Since the proposed MsSAN employs the multi-scale structure, it is difficult to evaluate its performance with different the SA blocks number. 
Here, we investigate the impact of the numbers of SA blocks in SAN on estimation performance, which is considered one of ablation experiments for our MsSAN.
As illustrated in Fig. \ref{fig:block_nmse}, the NMSE of SAN with different numbers of SA blocks are evaluated under different SNRs.
Regardless of the number of blocks in SAN, MsSAN consistently outperforms SAN in terms of NMSE.
This consistency highlights the effectiveness of the multi-scale approach cooperating with the SA module in improving channel estimation accuracy.

We evaluate the testing loss of the testing data as increasing training epochs and the validation losses of different networks are depicted as in Fig. \ref{fig:loss}.
As the number of training epochs increases, the loss of all models gradually decrease, indicating that the neural networks are continuously optimizing during the training process.
It is fact that both SAN and MsSAN start with lower loss values at the beginning of training phase, suggesting that they are capable of quickly extracting channel features.
Meanwhile, the faster convergence and lower loss of MsSAN demonstrate that the integration of multi-scale structure and the designed SA module result in more effective and accurate near-field channel estimation.

\section{Conclusion}\label{section:conclusion}
This paper proposed a multi-scale spatial-attention network (MsSAN) for the near-field MIMO channel estimation.
We first analyze and derive the spatial antenna correlation of near-field channels in the angular and distance domains separately.
The antenna correlation reveals that the inhomogeneity and complexity of the spatial features across the antenna elements.
Besides, the strong correlation between adjacent elements indicates the spatial correlation can be approximately described with subchannels.
Inspired by this, we employed the multi-scale architecture to refine different scale of subchannels.
In MsSAN, we designed the spatial-attention module composed by a spatial multi-head attention (SMA) and a gated spatial feed-forward network (GSFN) tailed to focus on the correlation of inter-subchannel features.
Specifically, the sum of dot products was introduced as spatial attention rather than cross-covariance to weight subchannel features at different scales.
Simulation results demonstrated that the proposed MsSAN outperforms other estimation schemes.
Ablation experiments show MsSAN incorporating a multi-scale feature extraction capability outperforms SAN with a single-scale structure.

%
%
%
%

\bibliography{ref}
\bibliographystyle{IEEEtran}
\end{document}